\documentclass[letterpaper,twocolumn,twoside]{IEEEtran} 
\newcommand{\figsize}{3.4in}

\usepackage{graphicx,psfrag}
\usepackage[cmex10]{amsmath}
\usepackage{amsfonts,amssymb,latexsym,bm}
\usepackage{hyperref} 
\usepackage{cite}
\usepackage{enumerate,subcaption}
\usepackage[usenames]{color}
\usepackage{stmaryrd}
\usepackage{footmisc}

\usepackage{algorithm,algpseudocode}

\algnewcommand{\Inputs}[1]{%
  \State \textbf{inputs:}
  \Statex \hspace*{\algorithmicindent}\parbox[t]{.8\linewidth}{\raggedright #1}
}
\algnewcommand{\Initialize}[1]{%
  \State \textbf{initialize:}
  \Statex \hspace*{\algorithmicindent}\parbox[t]{.8\linewidth}{\raggedright #1}
}

\graphicspath{{fig/}}

\usepackage{pbox}

\newcommand{\textb}[1]{\textcolor{black}{#1}}

\newcommand{\blue}{\color{black}}
\newcommand{\black}{\color{black}}

\renewcommand{\hat}{\widehat}

\newcommand{\defn}{\triangleq}
\newcommand{\mat}[1]{\ensuremath{\begin{bmatrix}#1\end{bmatrix}}}

 \newcommand{\mc}[1]{\ensuremath{\mathcal{#1}}}
\newcommand{\Real}{{\mathbb{R}}}

\newcommand{\tran}{^{\text{\textsf{T}}}}
\newcommand{\herm}{^{\text{\textsf{H}}}}
\newcommand*\dif{\mathop{}\!\mathrm{d}} 
\newcommand{\zero}{\mathbf{0}}
\newcommand{\one}{\mathbf{1}}

\DeclareMathOperator{\E}{\mathbb{E}}
\DeclareMathOperator{\Exp}{\mathbb{E}}
\DeclareMathOperator{\var}{var}
\DeclareMathOperator{\cov}{Cov}

\DeclareMathOperator{\tr}{tr}
\DeclareMathOperator{\diag}{diag}
\DeclareMathOperator{\Diag}{Diag}

\DeclareMathOperator{\vect}{vec}

\DeclareMathOperator*{\argmax}{arg\,max}

\renewcommand{\eqref}[1]{(\ref{eq:#1})}

\newcommand{\Figref}[1]{Figure~\ref{fig:#1}}
\newcommand{\figref}[1]{Fig.~\ref{fig:#1}}
\newcommand{\tabref}[1]{Table~\ref{tab:#1}}

\newcommand{\secref}[1]{Sec.~\ref{sec:#1}}

\renewcommand{\algref}[1]{Alg.~\ref{alg:#1}}
\newcommand{\lineref}[1]{line~\ref{line:#1}}

\newcommand{\textsbif}[1]{\textsf{\textbf{\textit{#1}}}}
\newcommand{\normal}{\mathcal{N}}
\newcommand{\pX}{p_{\textsbif{X}}}
\newcommand{\px}{p_{\textsbif{x}}}
\newcommand{\pY}{p_{\textsbif{Y}}}

\newcommand{\pYgX}{p_{\textsbif{Y}|\textsbif{X}}}
\newcommand{\pygx}{p_{\textsbif{y}|\textsbif{x}}}
\newcommand{\pXgY}{p_{\textsbif{X}|\textsbif{Y}}}
\newcommand{\pxgy}{p_{\textsbif{x}|\textsbif{y}}}
\newcommand{\pXY}{p_{\textsbif{X},\textsbif{Y}}}

\newcommand{\qt}{q^t}

\newcommand{\taup}{\tau+}

\newcommand{\Abf}{\bm{A}}
\newcommand{\Cbf}{\bm{C}}
\newcommand{\Dbf}{\bm{D}}

\newcommand{\Hbf}{\bm{H}}
\newcommand{\Ibf}{\bm{I}}
\newcommand{\Pbf}{\bm{P}}

\newcommand{\Sbf}{\bm{S}}
\newcommand{\Ubf}{\bm{U}}
\newcommand{\Vbf}{\bm{V}}
\newcommand{\Wbf}{\bm{W}}
\newcommand{\Xbf}{\bm{X}}
\newcommand{\Ybf}{\bm{Y}}

\newcommand{\bbf}{\bm{b}}
\newcommand{\cbf}{\bm{c}}
\newcommand{\ebf}{\bm{e}}

\newcommand{\hbf}{\bm{h}}
\newcommand{\gbf}{\bm{g}}

\newcommand{\rbf}{\bm{r}}

\newcommand{\wbf}{\bm{w}}
\newcommand{\xbf}{\bm{x}}
\newcommand{\ybf}{\bm{y}}

\newcommand{\Ab}{\bm{A}}
\newcommand{\Bb}{\bm{B}}
\newcommand{\Bbf}{\bm{B}}
\newcommand{\Cb}{\bm{C}}

\newcommand{\Hb}{\bm{H}}
\newcommand{\Lb}{\bm{L}}

\newcommand{\Rb}{\bm{R}}
\newcommand{\Sb}{\bm{S}}
\newcommand{\Pb}{\bm{P}}
\newcommand{\Ub}{\bm{U}}
\newcommand{\Vb}{\bm{V}}
\newcommand{\Wb}{\bm{W}}
\newcommand{\Xb}{\bm{X}}
\newcommand{\Yb}{\bm{Y}}
\newcommand{\Zb}{\bm{Z}}

\newcommand{\Ib}{\bm{I}}
\newcommand{\cb}{\bm{c}}
\newcommand{\eb}{\bm{e}}
\newcommand{\rb}{\bm{r}}
\renewcommand{\sb}{\bm{s}} 

\newcommand{\gb}{\bm{g}}
\newcommand{\wb}{\bm{w}}
\newcommand{\yb}{\bm{y}}
\newcommand{\zb}{\bm{z}}
\newcommand{\xb}{\bm{x}}
\newcommand{\xhat}{\hat{\bm{x}}}
\newcommand{\abf}{\bm{a}}

\newcommand{\thetabf}{\bm{\theta}}

\newcommand{\Thetabf}{\bm{\Theta}}
\newcommand{\Psibf}{\bm{\Psi}}

\newcommand{\gammabf}{\bm{\gamma}}
\newcommand{\betabf}{\bm{\beta}}
\newcommand{\rhobf}{\bm{\rho}}

\newcommand{\iter}{t}
\newcommand{\iters}{{t+1}}

\newcommand{\MMSE}{_{\text{\sf MMSE}}}
\newcommand{\ML}{_{\text{\sf ML}}}
\newcommand{\Dkl}{D_{\text{\sf KL}}}

\title{Bilinear Recovery using Adaptive Vector-AMP}
\author{Subrata Sarkar, 
        Alyson K. Fletcher, 
        Sundeep Rangan, 
        and 
        Philip Schniter\IEEEauthorrefmark{1} 
        \thanks{S. Sarkar (email: sarkar.51@osu.edu)
                and P. Schniter (email: schniter.1@osu.edu)
                are with 
                the Department of Electrical and Computer Engineering,
                The Ohio State University, Columbus, OH, USA, 43202.
                }%
        \thanks{A.~K.~Fletcher (email: akfletcher@ucla.edu) 
                is with
                the Department of Statistics and Electrical Engineering,
                the University of California, Los Angeles, CA, USA, 90095.
                }%
        \thanks{S. Rangan (email: srangan@nyu.edu) 
                is with
                the Department of Electrical and Computer Engineering,
                New York University, Brooklyn, NY, USA, 11201.
                }%
        \thanks{Please direct all correspondence to 
                Prof.\ Philip Schniter,
                Dept. ECE, 2015 Neil Ave., Columbus OH 43210,
                phone 614.247.6488, fax 614.292.7596.}%
        }

\begin{document}
\setlength{\arraycolsep}{0.5mm}
\maketitle

\begin{abstract}
We consider the problem of jointly recovering the vector $\boldsymbol{b}$ and the matrix $\boldsymbol{C}$ from noisy measurements $\boldsymbol{Y} = \boldsymbol{A}(\boldsymbol{b})\boldsymbol{C} + \boldsymbol{W}$, where $\boldsymbol{A}(\cdot)$ is a known affine linear function of $\boldsymbol{b}$ (i.e., $\boldsymbol{A}(\boldsymbol{b})=\boldsymbol{A}_0+\sum_{i=1}^Q b_i \boldsymbol{A}_i$ with known matrices $\boldsymbol{A}_i$).
This problem has applications in 
matrix completion, robust PCA, dictionary learning, 
self-calibration, blind deconvolution, joint-channel/symbol estimation,
compressive sensing with matrix uncertainty, and many other tasks.
To solve this bilinear recovery problem, we propose the Bilinear Adaptive Vector Approximate Message Passing (BAd-VAMP) algorithm.
We demonstrate numerically that the proposed approach is competitive with other state-of-the-art approaches to bilinear recovery, including lifted VAMP and Bilinear GAMP.
\end{abstract}


\section{Introduction} \label{sec:intro} 

\subsection{Motivation} \label{sec:motivation}

Many problems of interest in science and engineering can be formulated as estimation of a structured matrix $\Zb$ from noisy or incomplete measurements.
The type of structure in $\Zb$ determines the specific subproblem to be solved.

For example, when $\Zb$ has a low-rank structure and only a subset of its entries are observed, the problem is known as \emph{matrix completion} \cite{Candes:PROC:10}.
When $\Zb=\Lb+\Sb$ for low-rank $\Lb$ and sparse $\Sb$, the problem of estimating $\Lb$ and $\Sb$ is known as \emph{robust principle components analysis} (RPCA) \cite{Candes:JACM:11}.
When $\Zb=\Bb\Cb$ with sparse $\Cb$, the problem of estimating $\Bb$ and $\Cb$ is known as \emph{dictionary learning} \cite{Rubinstein:PROC:10}.
When $\Zb=\Bb\Cb$ with nonnegative $\Bb$ and $\Cb$, the problem is known as \emph{nonnegative matrix factorization} (NMF) \cite{Lee:NIPS:01}.

Sometimes $\Zb$ has a more complicated structure.
For example, the problems of \emph{self-calibration} and \emph{blind (circular) deconvolution} \cite{Ling:IP:15} can be formulated using $\Zb=\Diag(\Hb \bbf) \Psibf\Cb$,
where $\Hb$ and $\Psibf$ are known and $\bbf$ and $\Cb$ are to be estimated.\footnote{Here and in the sequel, we use lowercase bold notation for vectors and uppercase bold notation for matrices.}
The problem of \emph{compressive sensing (CS) with matrix uncertainty} \cite{Zhu:TSP:11} can be formulated using $\zb=\sum_i b_i \Ab_i \cb$, where $\{\Ab_i\}$ are known and where $\bbf$ and sparse $\cb$ are to be estimated.
The latter covers the problem of \emph{joint channel-symbol estimation} \cite{Kaleh:TCOM:94}, in which case $b_i$ are the data symbols, $\cb$ contains (possibly sparse) channel coefficients, and the known $\{\Ab_i\}$ are determined by the modulation scheme.
The more general problem of \emph{matrix CS} \cite{Candes:TIT:11,Waters:NIPS:11} results from
\begin{align}
z_m = \tr\{\Ab_m\tran(\Lb+\Sb)\} \text{~for~} m=1,\dots,M,
\end{align}
where $\{\Ab_m\}$ are known and the goal is to estimate low-rank $\Lb$ and sparse $\Sb$.

\subsection{Prior Work} \label{sec:prior}

Many algorithms have been developed to solve the above problems. 
Some solve a convex relaxation of the original problem, while others attack non-convex formulations via alternating methods, greedy methods, variational methods, message-passing methods, and other techniques.

For matrix completion, well-known approaches include 
the nuclear-norm-based convex optimization method IALM \cite{Lin:10}, 
the non-convex successive over-relaxation approach LMAFit \cite{Wen:MPC:12},
the Grassmanian gradient-descent approach GROUSE \cite{Balzano:ALL:10}, 
the greedy hard-thresholding approach Matrix-ALPS \cite{Kyrillidis:JMIV:14}, 
and the variational-Bayes method VSBL \cite{Babacan:TSP:12}. 
For RPCA, there are also versions of
IALM \cite{Lin:10}, LMaFit \cite{Wen:MPC:12}, and VSBL \cite{Babacan:TSP:12}, as well as a robust cousin of GROUSE, called GRASTA \cite{He:CVPR:12}.
For dictionary learning, there is the 
greedy K-SVD algorithm \cite{Aharon:TSP:06},
the online SPAMS approach \cite{Mairal:JMLR:10},
and the ER-SpUD approach from \cite{Spielman:COLT:12}.
A unified approach to matrix completion, RPCA, and dictionary learning was proposed in \cite{Parker:TSP:14a,Parker:TSP:14b,Kabashima:TIT:16} using an extension of the approximate message-passing (AMP) methodology from \cite{Donoho:PNAS:09,Rangan:ISIT:11}.
The resulting ``bilinear generalized AMP'' (BiGAMP) algorithm was compared to the aforementioned methods in \cite{Parker:TSP:14b} and found (empirically) to be competitive, if not superior, in phase transition and runtime.
A related approach known as LowRAMP was proposed \cite{Matsushita:NIPS:13} and analyzed in \cite{Lesieur:JSM:17,Miolane:17}.

For self-calibration and blind deconvolution, well-known approaches include
the convex relaxations from \cite{Bilen:TSP:14, 
Ahmed:TIT:14,Ling:IP:15} 
and the alternating method from \cite{Hedge:TSP:11}. 
For CS with matrix uncertainty, there is the award-winning non-convex method \cite{Zhu:TSP:11}.
For matrix CS, the well-known papers \cite{Candes:TIT:11,Agarwal:AS:12,Wright:II:13} proposed convex approaches and \cite{Waters:NIPS:11,Kyrillidis:JMIV:14} proposed greedy approaches.
See the recent overview \cite{Davenport:JSTSP:16} for many other works.
An AMP-based approach to self-calibration, blind deconvolution, CS with matrix uncertainty, and matrix CS was proposed in \cite{Parker:JSTSP:16} and analyzed in \cite{Schulke:PRE:16}.
This ``parametric BiGAMP'' (PBiGAMP) was compared to the above works in \cite{Parker:JSTSP:16} and found to yield improved empirical phase transitions.

More recently, AMP methods for bilinear inference were proposed using the ``lifting'' approach (see, e.g., \cite{Candes:CPAM:13,Ahmed:TIT:14,Ling:IP:15,Davenport:JSTSP:16} for seminal papers on lifting).
To illustrate the idea, suppose that the measurement vector $\ybf\in\Real^M$ takes the form
\begin{align}
\ybf 
&= \sum_{i=1}^Q \sum_{j=1}^N b_i \abf_{i,j} c_j + \wbf  
\label{eq:bilinear} ,
\end{align}
where \textb{$\abf_{i,j}\in\Real^M$ is known for all $i,j$} and the goal is to recover $\bbf=[b_1,\dots,b_Q]\tran$ and $\cbf=[c_1,\dots,c_N]\tran$ in the presence of white noise $\wbf$.
Rewriting the measurements as
\begin{align}
\ybf 
&= \sum_{i=1}^Q b_i \underbrace{\begin{bmatrix}\abf_{i,1},\cdots,\abf_{i,N}\end{bmatrix}}_{\displaystyle \defn \Abf_{i}} \cbf + \wbf 
\label{eq:bilinear2} \\
&= \underbrace{\begin{bmatrix}\Abf_{1}\cdots\Abf_{Q}\end{bmatrix}}_{\displaystyle \defn \Abf} \begin{bmatrix}b_1\cbf\\\vdots\\b_Q\cbf\end{bmatrix} + \wbf \\
&= \Abf\xbf + \wbf \text{~~for~~} \xbf= \bbf\otimes\cbf = \vect(\cbf\bbf\tran) 
\label{eq:linear} ,
\end{align}
we see that the noisy bilinear recovery problem \eqref{bilinear} can be rewritten as the noisy linear recovery problem \eqref{linear} with a rank-one structure on (the matrix form of) $\xbf$. 
Thus, if this low-rank signal structure can be exploited by a linear inference algorithm, then bilinear inference can be accomplished.
This is precisely what was proposed in \cite{Romanov:PNAS:18}, building on the non-separable-denoising version of the AMP algorithm from \cite{Metzler:TIT:16}. 
A rigorous analysis of ``lifted AMP'' was presented in \cite{Berthier:17}.

The trouble with AMP is that its behavior is understood only in the case of \textb{large \cite{Rush:ISIT:16} or infinitely} large, i.i.d.\ (sub) Gaussian $\Abf$ \cite{Bayati:TIT:11,Bayati:AAP:15}.
Even small deviations from this scenario (e.g., mildly ill-conditioned and/or non-zero-mean $\Abf$) can cause AMP to diverge \cite{Rangan:TIT:16a,Caltagirone:ISIT:14,Vila:ICASSP:15}.
To address this issue, an alternative called Vector AMP (VAMP) was proposed and analyzed in \cite{Rangan:VAMP}, with close connections to expectation propagation \cite{Opper:NIPS:05} (see also \cite{Fletcher:ISIT:16,Ma:IA:17,Takeuchi:ISIT:17}).
There it was established that, if $\Abf$ is an infinitely large right-rotationally invariant\footnote{If $\Abf$ is right-rotationally invariant then its singular value decomposition $\Abf=\Ubf\Sbf\Vbf\tran$ has Haar distributed $\Vbf$, i.e., $\Vbf$ is  uniformly distributed over the group of orthogonal matrices.} random matrix and the denoising function used by VAMP is separable and Lipschitz, then VAMP's performance can be exactly predicted by a scalar state-evolution that also provides testable conditions for optimality. 
Since the class of right-rotationally invariant matrices is much larger than the class of i.i.d.\ Gaussian matrices, VAMP is much more robust than AMP with regards to the construction of $\Abf$.
For example, VAMP has no problem with ill-conditioned or mean-shifted matrices \cite{Rangan:VAMP}.

Very recently, \cite{Fletcher:NIPS:18} performed a rigorous analysis of VAMP under \emph{non}-separable Lipschitz denoisers, showing that---here too---VAMP's behavior is exactly predicted by a scalar state-evolution when $\Abf$ is \textb{infinitely} large and right-rotationally invariant.
Furthermore, \cite{Fletcher:NIPS:18} demonstrated the success of lifted VAMP on bilinear problems such as self-calibration and CS with matrix uncertainty.
In addition, \cite{Fletcher:NIPS:18} gave evidence that, like AMP, the PBiGAMP algorithm is sensitive to deviations from the i.i.d.\ assumptions used in its derivation \cite{Parker:JSTSP:16} and analysis \cite{Schulke:PRE:16}.
For this reason, lifted VAMP significantly outperformed PBiGAMP in some cases \cite{Fletcher:NIPS:18}.

Despite its good performance and rigorous analyses under \textb{infinitely} large right-rotationally invariant random $\Abf$, 
lifted VAMP suffers from computational issues brought on by the lifting itself:
The $N+Q$ unknowns $[\bbf,\cbf]$ in the bilinear problem \eqref{bilinear} manifest as $NQ$ unknowns $\xbf$ after lifting to \eqref{linear}.  
This is a serious problem when $N$ and $Q$ are both large.
As a concrete example, consider the application of lifting to (square) dictionary learning, where the goal is to recover $\Bb\in\Real^{N\times N}$ and sparse $\Cb\in\Real^{N\times L}$ from noisy measurements $\Ybf=\Bb\Cb+\Wb$.
This bilinear relationship can be lifted via 
\begin{align}
\Yb
&= \sum_{ij} b_{i,j} \Ab_{i,j} \Cb + \Wbf \\
&= \underbrace{[\Ab_{1,1} \cdots \Ab_{N,N}]}_{\displaystyle \defn \Abf\in\Real^{N\times N^3}} 
  \hspace{-10mm}
  \underbrace{\big(\bbf \otimes \Cb \big)}_{\displaystyle \hspace{12mm} \defn \Xbf \in\Real^{N^3\times L}} \hspace{-10mm} +\, \Wbf,
\end{align}
where $\Ab_{i,j}\in\Real^{N\times N}$ is constructed with a $1$ in the $(i,j)$th position and zeros elsewhere,
and where $\bbf=[b_{1,1},\dots,b_{N,N}]\tran\in\Real^{N^2}$.
Even at the relatively small patch size of $8\times 8$ (i.e., $N=64$), the matrix $\Abf$ has dimension $64\times 262\,144$, and the unknown matrix $\Xbf$ has dimension $262\,144\times L$.
The rule-of-thumb $L=5 N\ln N$ \cite{Spielman:COLT:12} then gives $L=1331$, in which case $\Xbf$ contains $3.5\times 10^8$ entries, which leads to difficulties with computation and memory.

\subsection{Contributions}

In this paper, we present a novel VAMP-based approach to bilinear recovery.
With the aim of computational efficiency, we avoid lifting and instead build on the recently proposed Adaptive VAMP framework from \cite{Fletcher:NIPS:17}.
However, different from \cite{Fletcher:NIPS:17}, which focused on noisy \emph{linear} recovery, we focus on noisy \emph{bilinear} recovery. 

In particular, we focus on recovering $\{b_i\}$ and $\Cb$ from noisy measurements $\Yb\in\Real^{M\times L}$ of the form
\begin{align}
\Yb &= \sum_{i=1}^Q b_i \Ab_i \Cbf + \Wbf 
\label{eq:bilinear3},
\end{align}
where $\{\Ab_i\}$ are known and $\Wbf$ contains white noise. 
Note that \eqref{bilinear3} is a multiple-measurement vector (MMV) extension of \eqref{bilinear2}, and that it covers all of the motivating problems discussed in \secref{motivation}. 
\textb{For example, in self-calibration, where we estimate $\bbf$ and $\Cbf$ from $\Ybf=\Diag(\Hbf\bbf)\Psibf\Cbf+\Wbf$, we can set $\Abf_i=\Diag(\hbf_i)\Psibf$, where $\hbf_i$ is the $i$th column of $\Hbf$.
Or, in dictionary learning, where we estimate $\Bbf$ and $\Cbf$ from $\Ybf=\Bbf\Cbf+\Wbf$, we can write $\Bbf=\sum_{i=1}^{MN} b_i \Abf_i$ for $\Abf_i=\ebf_{\langle i-1\rangle_M}\ebf_{\lfloor (i-1)/M \rfloor}\tran$, where $\langle i\rangle_M$ denotes $i$-modulo-$M$, $\lfloor \cdot \rfloor$ denotes floor, and $\{\ebf_i\}$ is the standard basis.}

When deriving\footnote{Although the derivation treats the entries of $\Cbf$ as random variables and the associated denoiser as Bayesian, the final algorithm is more general in that it only requires the denoiser to be Lipschitz.} the proposed method, we treat $\{b_i\}$ as deterministic unknowns and the entries of $\Cbf$ as random variables.
The prior distribution on $\Cbf$ is assumed to be known up to some (possibly) unknown hyperparameters, which are learned jointly with $\{b_i\}$ and $\Cbf$.
Also, $\Wbf$ is treated as additive white Gaussian noise (AWGN) with an unknown variance that is also learned. More details are provided in the sequel.

We show (empirically) that the proposed Bilinear Adaptive VAMP (BAd-VAMP) method performs as well as the EM-PBiGAMP algorithm from \cite{Parker:JSTSP:16}, with regard to accuracy and computational complexity, when the underlying matrices are i.i.d., as assumed for the derivation of PBiGAMP.
However, we will show that BAd-VAMP outperforms EM-PBiGAMP when the underlying matrices become ill-conditioned.
In the ill-conditioned case, we show that BAd-VAMP performs as well as, and sometimes significantly better than, lifted VAMP.
However, BAd-VAMP is much more computationally efficient due to its avoidance of lifting.
In this sense, the proposed BAd-VAMP is shown to be accurate, robust, and computationally efficient.

\textit{Notation:}
In this paper, we use 
boldface uppercase letters to denote matrices (e.g., $\Xbf$), boldface lowercase letters for vectors (e.g., $\xbf$) and non-bold letters for scalars (e.g., $x$). 
Given a matrix $\Xb$, we use $\xb_l$ to denote the $l$th column and $x_{nl}$ to denote the element in the $n$th row and $l$th column. 
We use $\Exp[f(\xbf)|b]$ to denote the expectation of $f(\xbf)$ w.r.t. the density $b$, i.e., $\Exp[f(\xbf)|b]=\int f(\xb)b(\xb)\dif\xb$, and we use $\var[f(\xbf)|b]$ for the corresponding variance. 
\textb{We use $\Diag(\xbf)$ to denote the diagonal matrix created from vector $\xbf$, and $\diag(\Xbf)$ to denote the vector of elements on the diagonal of the matrix $\Xbf$.}   

\section{Proposed Framework}

In an effort to make our algorithmic development more consistent with the VAMP papers \cite{Rangan:VAMP,Fletcher:NIPS:17,Fletcher:NIPS:18}, we now make some minor notational changes relative to \eqref{bilinear3}.
First, we will use the notation $\Abf(\bbf) \defn \sum_i b_i \Abf_i$ to be concise. 
Second, the quantities $b_i$ and $\Cbf$ in \eqref{bilinear3} will be changed to $\theta_{A,i}$ and $\Xbf$, respectively.

The problem of interest can thus be stated as follows: 
estimate the matrix $\Xbf \in \Real^{N\times L}$ and learn the parameters $\Thetabf\defn\{\thetabf_A,\thetabf_x,\gamma_w\}$ in the statistical model 
\begin{subequations}\label{eq:model}
\begin{align}
\Yb &= \Ab(\thetabf_A)\Xbf + \Wbf 
\label{eq:Y}\\
\Xbf &\sim \pX(\cdot;\thetabf_x),\quad w_{ml} \stackrel{\text{i.i.d.}}{\sim}\normal(0,\gamma_w^{-1}) ,
\end{align}
\end{subequations}
where $\Abf(\cdot)$ is a known matrix-valued linear function, and
$\pX(\cdot;\thetabf_x)$ is a density on $\Xbf$ parameterized by the vector $\thetabf_x$.
Here, $\gamma_w$ is the noise precision, i.e., the inverse noise variance.

More precisely, we aim to compute the maximimum-likelihood (ML) estimate of $\Thetabf$ and, under that estimate, compute the minimum mean-squared error (MMSE) estimate of $\Xbf$, i.e.,
\begin{align}
\hat{\Thetabf}\ML 
&= \argmax_{\Thetabf} \pY(\Yb;\Thetabf)
\label{eq:ThetaML} \\
\hat{\Xb}\MMSE   
&= \Exp[\Xbf|\Yb;\hat{\Thetabf}\ML]
\label{eq:Xmmse} .
\end{align}
In \eqref{ThetaML}, $\pY(\Yb;\Thetabf)$ is the likelihood function of $\Thetabf$, which can be written as 
\begin{align}
\pY(\Yb;\Thetabf) 
&= \int \pX(\Xb;\Thetabf)\pYgX(\Yb|\Xb;\Thetabf)\dif\Xb
\label{eq:pYTheta} .
\end{align}
In \eqref{Xmmse},
the expectation is taken over the posterior density
\begin{align}
\pXgY(\Xb|\Yb;\hat{\Thetabf}\ML)
&= \frac{\pX(\Xb;\hat{\Thetabf}\ML)\pYgX(\Yb|\Xb;\hat{\Thetabf}\ML)}{\pY(\Yb;\hat{\Thetabf}\ML)} .
\end{align}

The statistical model \eqref{model} implies that\footnote{In \eqref{pY|X}-\eqref{pX}, to promote notational simplicity, the left side of the equation is written using $\Thetabf$ even though the right side depends on a subset of $\Thetabf$.}
\begin{align}
\pYgX(\Yb|\Xb;\Thetabf) 
&= \prod_{l=1}^L \pygx(\yb_l|\xb_l;\Thetabf) 
\label{eq:pY|X}
\end{align}
where
\begin{align}
\pygx(\yb|\xb;\Thetabf)
&= \normal\big(\yb; \Ab(\thetabf_A)\xb,\Ib/\gamma_w\big) 
\label{eq:py|x}.
\end{align}
For simplicity, we treat $\{\xbf_l\}_{l=1}^L$ as i.i.d.\ in the sequel, so that
\begin{align}
\pX(\Xb;\Thetabf) 
&= \prod_{l=1}^L \px(\xb_l;\thetabf_x) 
\label{eq:pX}  
\end{align}
for some density $\px(\cdot;\thetabf_x)$ parameterized by $\thetabf_x$.
In this case, the posterior density decouples as  
\begin{align}
\pXgY(\Xb|\Yb;\Thetabf)
&\propto \prod_{l=1}^L \px(\xb_l;\Thetabf)\pygx(\yb_l|\xb_l;\Thetabf)
\label{eq:pX|YTheta} .
\end{align}

\section{Background} \label{sec:background}

\subsection{Background on VAMP} \label{sec:VAMP}

Recalling \eqref{Xmmse}, we are interested in computing the MMSE estimate of $\Xbf$ from the noisy measurements $\Ybf$. 
This problem is solved (under certain conditions) by the VAMP approach from \cite{Rangan:VAMP}.
We now review VAMP at the detail needed for further development of BAd-VAMP.

From \eqref{Xmmse}, the MMSE estimate of $\Xbf$ equals the mean of the posterior pdf $\pXgY$. 
Because $\pXgY$ decouples across the columns of $\Xbf$, as in \eqref{pX|YTheta}, it suffices to consider a single column and drop the $l$ notation for simplicity.
Also, for now, we will assume that $\Thetabf$ are the true parameters used to generate $(\Ybf,\Xbf)$ and drop the $\Thetabf$ notation for simplicity;
we will revisit the estimation of $\Thetabf$ in \secref{EM}.
With these simplifications, \eqref{model} reduces to 
\begin{align}
\yb = \Ab\xbf + \normal(\zero,\Ib/\gamma_w),\quad \xbf\sim\px
\label{eq:regression_prob} .
\end{align}

Recall that the MMSE estimate of $\xbf$ equals 
\begin{align}
\E[\xbf|\ybf] &= \int \xb \, \pxgy(\xb|\yb) \dif\xb 
\label{eq:Ex|y}
\end{align}
for the posterior density
\begin{align}
\pxgy(\xb|\yb) &= Z(\ybf)^{-1} \px(\xb)\pygx(\yb|\xb) 
\label{eq:px|y},
\end{align}
where $Z(\ybf)$ is the normalizing constant 
\begin{align}
Z(\ybf) &\defn \int\px(\xb)\pygx(\yb|\xb)\dif\xb 
\label{eq:Z}.
\end{align}
For high-dimensional $\xbf$, the integrals in \eqref{Ex|y} and \eqref{Z} are difficult to compute directly.
Thus other methods must be used.

Variational inference (VI) \cite{Wainwright:FTML:08} 
can be used to bypass the computation of $Z(\ybf)$.
For example, notice that the true posterior $\pxgy$ can be recovered by solving the variational optimization (over densities)
\begin{align}
\hat{q} &= \arg\min_{q} \Dkl(q\,\|\,\pxgy)
\label{eq:VI} ,
\end{align}
where $\Dkl(q\,\|\,p)$ denotes the KL divergence from $p$ to $q$, i.e.,
\begin{align}
\Dkl(q\,\|\,p) &\defn \int q(\xb) \ln \frac{q(\xb)}{p(\xb)} \dif\xb 
\label{eq:Dkl} .
\end{align}
Plugging \eqref{px|y} into \eqref{Dkl}, we see that
\begin{align}
\Dkl(q\,\|\,\pxgy) &= \Dkl(q\,\|\,\px) + \Dkl(q\,\|\,\pygx) + H(q) 
\nonumber\\&\quad
+ \ln Z(\ybf)
\label{eq:Dkl2}
\end{align}
where $H(q) \defn -\int q(\xb)\ln q(\xb)\dif\xb$ is the differential entropy of $\xbf\sim q$.
Thus it follows from \eqref{VI} and \eqref{Dkl2} that
\begin{align}
\hat{q} &= \arg\min_{q} \big\{ \Dkl(q\,\|\,\px) + \Dkl(q\,\|\,\pygx) + H(q) \big\}  
\label{eq:VI2} ,
\end{align}
which bypasses $Z(\ybf)$.
Still, solving \eqref{VI2} is difficult in most cases of interest.
The typical response is to impose constraints on $q$, but doing so compromises $\hat{q}$ and its mean.

We take a different approach.
Using the ``Gibbs free energy''
\begin{align}
J(q_1,q_2,q_3)\defn \Dkl(q_1\|\px) + \Dkl(q_2\|\pygx) + H(q_3) ,
\end{align}
one can rewrite \eqref{VI2} as\footnote{\textb{We minimize over $q_1$ and $q_2$ because $\Dkl(q_1\|\px)$ and $\Dkl(q_2\|\pygx)$ are convex, while we maximize over $q_3$ because $H(q_3)$ is concave.}}
\begin{subequations} \label{eq:J_exact}
\begin{align}
&\arg\min_{q_1} \min_{q_2} \max_{q_3} J(q_1,q_2,q_3)\\
&\;\text{s.t. }  q_1 = q_2 = q_3 \label{eq:bequal} .
\end{align}
\end{subequations}
But, as discussed earlier, \eqref{J_exact} is difficult to solve.
In the \emph{expectation consistent approximate inference} (EC) scheme proposed by Opper and Winther in \cite{Opper:NIPS:05}, the density constraint \eqref{bequal} is relaxed to moment-matching constraints, i.e.,
\begin{subequations} \label{eq:EC_problem}
\begin{align}
&\arg \min_{q_1} \min_{q_2} \max_{q_3} J(q_1,q_2,q_3)\\
&\; \text{s.t. } \Exp[\xbf|q_1] = \Exp[\xbf|q_2] = \Exp[\xbf|q_3] \label{eq:moment1}\\
&\quad \tr\{\cov[\xbf|q_1]\} = \tr\{\cov[\xbf|q_2]\} = \tr\{\cov[\xbf|q_3]\}\label{eq:moment2} ,
\end{align}
\end{subequations}
where $\Exp[\xbf|\qt]$ denotes $\Exp[\xbf]$ under $\xbf\sim\qt$.
This yields stationary points of the form 
\begin{subequations} \label{eq:EC_stationary_points}
\begin{align}
q_1(\xb) &\propto \px(\xb)\exp\left(-\tfrac{\gamma_1}{2}\|\xb-\rb_1\|_2^2\right) \label{eq:b1}\\
q_2(\xb) &\propto \pygx(\yb|\xb)\exp\left(-\tfrac{\gamma_2}{2}\|\xb-\rb_2\|_2^2\right) \\
q_3(\xb) &\propto \exp\left(-\tfrac{\eta}{2}\|\xb-\xhat\|_2^2\right) ,
\end{align} 
\end{subequations}
for $\{\rb_1,\gamma_1,\rb_2,\gamma_2,\xhat,\eta\}$ that lead to satisfaction of \eqref{moment1}-\eqref{moment2}.
Various approaches can be used to solve for $\{\rb_1,\gamma_1,\rb_2,\gamma_2,\xhat,\eta\}$. 
One is to 
alternate the update of $\{(\rb_1,\gamma_1)$, $(\xhat,\eta)\}$ and $\{(\rb_2,\gamma_2)$, $(\xhat,\eta)\}$ such that, at each iteration, the moments of $q_3$ are consistent with either $q_1$ or $q_2$. 
This approach is summarized in \algref{EP} 
using\footnote{In \cite{Fletcher:ISIT:16}, different $\gb_1$ and $\gb_2$ were proposed so that the EP algorithm accomplishes joint MAP estimation of $\xbf$ from $\ybf$, i.e., $\xhat=\arg\max_{\xbf} p(\xbf|\ybf)$.}
\begin{align}
\gb_1(\rb_1,\gamma_1) 
&\defn \frac{\int \xb\,\px(\xb)\normal(\xb;\rb_1,\Ib/\gamma_1)\dif\xb}{\int \px(\xb)\normal(\xb;\rb_1,\Ib/\gamma_1)\dif\xb}
\label{eq:g1_defn} \\
\gb_2(\rb_2,\gamma_2) 
&\defn \frac{\int \xb\,\pygx(\yb|\xb)\normal(\xb;\rb_2,\Ib/\gamma_2)\dif\xb}{\int \pygx(\yb|\xb)\normal(\xb;\rb_2,\Ib/\gamma_2)\dif\xb}
\label{eq:g2_defn} ,
\end{align}
which, \textb{under these definitions of $\gb_1$ and $\gb_2$, can be recognized as} an instance of \emph{expectation propagation} (EP)~\textb{\cite[Sec. 3.2]{Minka:Diss:01},\cite{Seeger:Tech:05,Heskes:UAI:02}}. 
In lines~\ref{line:eta1_iter_EP} and \ref{line:eta2_iter_EP},
$\gb_i'(\rb_i,\gamma_i)\in\Real^N$ denotes the diagonal of the Jacobian matrix of $\gb_i(\cdot,\gamma_i)$ at $\rb_i$, i.e., 
\begin{align}
\gb_i'(\rb_i,\gamma_i)
&\defn\diag\left(\frac{\partial\gb_i(\rb_i,\gamma_i)}{\partial \rb_i}\right),
\end{align}
and $\langle\xb\rangle$ denotes the average coefficient value, i.e., $\langle\xb\rangle\defn\frac{1}{N}\sum_{i=1}^N x_i$ for $\xb\in\Real^N$. 
Due to the form of $\pygx$ in \eqref{py|x}, it can be shown that 
\begin{eqnarray}
\gb_2(\rb_2,\gamma_2) 
&=& (\gamma_2\Ib + \gamma_w\Ab\tran\Ab)^{-1}(\gamma_2\rb_2+\gamma_w\Ab\tran\yb)
\label{eq:g2_slm} \quad\\
\langle \gb'_2(\rb_2,\gamma_2) \rangle
&=& \gamma_2 \tr\big\{(\gamma_2\Ib + \gamma_w\Ab\tran\Ab)^{-1} \big\}/N .
\end{eqnarray}
Meanwhile, the form of $\gb_1(\cdot)$ depends on $\px$ through \eqref{g1_defn}.

\begin{algorithm}[t]
\caption{VAMP algorithm \cite{Rangan:VAMP}}
\label{alg:EP}
\begin{algorithmic}[1]
\Initialize {$\rb_{1}^0,\gamma_{1}^0$}
\For{$t=0,\dots,T_{\max}$}
\State ${\xb}_{1}^{\iter}=\gb_1(\rb_{1}^{\iter},{\gamma}_{1}^{\iter})$\label{line:x1_iter_EP}
\State $1/{\eta}_{1}^{\iter}= \langle\gb_1'(\rb_{1}^{\iter},\gamma_{1}^{\iter})\rangle/\gamma_{1}^{\iter}$\label{line:eta1_iter_EP}
\State ${\gamma_{2}^{\iter}=\eta_{1}^{\iter}-\gamma_{1}^{\iter}}$
\State $\rb_{2}^{\iter}=(\eta_{1}^{\iter}\xb_{1}^{\iter}-\gamma_{1}^{\iter}\rb_{1}^{\iter})/\gamma_{2}^{\iter}$ 
\State ${\xb}_{2}^{\iter}=\gb_2(\rb_{2}^{\iter},{\gamma}_{2}^{\iter})$\label{line:x2_iter_EP}
\State $1/{\eta}_{2}^{\iter}= \langle\gb_2'(\rb_{2}^{\iter},\gamma_{2}^{\iter})\rangle/\gamma_{2}^{\iter}$\label{line:eta2_iter_EP}
\State $\gamma_{1}^{\iters}=\eta_{2}^{\iter}-\gamma_{2}^{\iter}$ 
\State $\rb_{1}^{\iters}=(\eta_{2}^{\iter}\xb_{2}^{\iter}-\gamma_{2}^\iter\rb_{2}^\iter) /(\eta_{2}^{\iter}-\gamma_{2}^{\iter})$
\EndFor 
\end{algorithmic}
\end{algorithm}

Based on the description above, one might wonder whether the EC stationary point
$\xhat = \Exp[\xbf|q_1] = \Exp[\xbf|q_2] = \Exp[\xbf|q_3]$
is a good approximation of the true conditional mean $\Exp[\xbf|\ybf]$,
and additionally one might question whether \algref{EP} converges to this $\xhat$.
Both of these concerns were resolved in the VAMP paper \cite{Rangan:VAMP}.
In particular, \cite{Rangan:VAMP} showed that,
when $\Ab$ is right rotationally invariant and asymptotically large,
the per-iteration behavior of \algref{EP} with $\gb_2(\cdot)$ from \eqref{g2_slm} and Lipschitz\footnote{While the original VAMP paper \cite{Rangan:VAMP} focused on separable Lipschitz $\gb_1(\cdot)$, \cite{Fletcher:NIPS:18} extended the results to non-separable Lipschitz $\gb_1(\cdot)$.}
$\gb_1(\cdot)$ is exactly predicted by a scalar state evolution.
Furthermore, in the case where $\gb_1(\cdot)$ is matched to generating $\px$ from \eqref{regression_prob} as in \eqref{g1_defn}, and where $\gb_2(\cdot)$ uses the true AWGN precision $\gamma_w<\infty$ as in \eqref{g2_defn}, the MSE of the fixed point $\xhat$ of \algref{EP} was shown in \cite{Rangan:VAMP} to match the MMSE predicted by the replica method \cite{Tulino:TIT:13}. 
\textb{This replica prediction is conjectured to be correct \cite{Reeves:ALL:17}, in which case the $\xhat$ generated by \algref{EP} under \eqref{g1_defn} and \eqref{g2_slm} will be MMSE for infinitely large, right-rotationally invariant $\Abf$ when the state evolution has unique fixed points. 
Note that, for infinitely large i.i.d.\ $\Abf$, the replica prediction has been proven to be correct \cite{Reeves:ISIT:16,Barbier:ALL:16}.}  

In the sequel, we will refer to \algref{EP} with generic Lipschitz $\gb_1(\cdot)$ as the VAMP algorithm, noting that it coincides with EP in the special case of Bayesian $\gb_1(\cdot)$ from \eqref{g1_defn}.
\textb{VAMP is more general than EP because it can be used with denoisers $\gb_1(\cdot)$ that have no probabilistic interpretation and still lead to precisely predictable behavior under infinitely large, right-rotationally $\Abf$ \cite{Rangan:VAMP,Fletcher:NIPS:18}.}
We note that, when VAMP is applied to the MMV model \eqref{Y}, a separate copy of $\{\rb_1,\gamma_1,\rb_2,\gamma_2,\xhat,\eta\}$ must be tracked for each column of $\Yb$.

\subsection{Background on Expectation Maximization} \label{sec:EM}

We now return to the case where $\Thetabf$ is unknown and the goal is to compute its ML estimate, $\hat{\Thetabf}\ML$.
From \eqref{ThetaML} and \eqref{pYTheta}, we have 
\begin{align}
\hat{\Thetabf}\ML 
&= \arg\min_{\Thetabf} -\ln \int \pX(\Xb;\Thetabf)\pYgX(\Yb|\Xb;\Thetabf)\dif\Xb
\label{eq:ML_cost} ,
\end{align}
but \eqref{ML_cost} is impractical to optimize directly due to the high dimensional integral.

Expectation-maximization (EM)~\cite{Dempster:JRSS:77} is a well known iterative approach to ML that alternates between i) minimizing an upper-bound of the negative log-likelihood and ii) tightening the upper-bound.
The EM algorithm is usually written as
\begin{subequations} \label{eq:EM_classical}
\begin{align}
Q(\Thetabf;\hat{\Thetabf}^\iter) 
&\defn -\Exp\big[\ln \pXY(\Xbf,\Yb;\Thetabf)\,\big|\, \Ybf;
        \hat{\Thetabf}^\iter\big] \\
\hat{\Thetabf}^\iters 
&= \arg\min_{\Thetabf} Q(\Thetabf;\hat{\Thetabf}^\iter) .
\end{align}
\end{subequations}
\textb{Letting $\qt = \pXgY(\cdot|\Yb;\hat{\Thetabf}^\iter)$, we can write 
\begin{subequations}
\begin{align}
Q(\Thetabf;\hat{\Thetabf}^\iter)
&= -\Exp\big[\ln \pX(\Xbf;\Thetabf)\,\big|\,\Ybf;\hat{\Thetabf}^\iter\big]
\nonumber\\&\quad
-\Exp\big[\ln \pYgX(\Ybf|\Xbf;\Thetabf)\,\big|\,\Ybf;\hat{\Thetabf}^\iter\big]\\
&=-\Exp\big[\ln \pX(\Xbf;\Thetabf)\,\big|\,\qt\big]
\nonumber\\&\quad
-\Exp\big[\ln \pYgX(\Ybf|\Xbf;\Thetabf)\,\big|\,\qt]\\
&= J(\qt,\qt,\qt;\Thetabf) + \text{const.}
\end{align}
\end{subequations}
where $J$, also known as the Gibbs free energy, is defined as}
\begin{align}
J(q_1,q_2,q_3;\Thetabf)
&\defn \Dkl(q_1\|\pX(\cdot,\Thetabf)) 
\nonumber\\&\quad
        + \Dkl(q_2\|\pYgX(\Ybf|\cdot;\Thetabf)) 
        + H(q_3) 
\label{eq:J_cost} .
\end{align}
\textb{Thus, \eqref{EM_classical} can} also be written as \cite{Neal:Jordan:98}
\begin{subequations} \label{eq:EM2}
\begin{align}
\qt 
&= \pXgY(\cdot|\Yb;\hat{\Thetabf}^\iter)
\label{eq:E_step}\\
\hat{\Thetabf}^\iters 
&= \arg\min_{\Thetabf} J(\qt,\qt,\qt;\Thetabf).
\label{eq:M_step} 
\end{align}
\end{subequations}
Note that $J(\qt,\qt,\qt;\Thetabf)$ is an upper bound on $-\ln \pY(\Yb;\Thetabf)$ for \emph{any} $\qt$ since 
\begin{align}
J(\qt,\qt,\qt;\Thetabf)
&= -\ln \pY(\Yb;\Thetabf) + \Dkl(\qt\,\|\,\pYgX(\Ybf|\cdot;\Thetabf)) ,
\end{align}
where $\Dkl\geq 0$ by construction.
Thus, while the specific choice of $\qt$ in \eqref{E_step} yields a tight upper bound in that
\begin{align}
J(\qt,\qt,\qt;\hat{\Thetabf}^\iter) = -\ln \pY(\Yb;\hat{\Thetabf}^\iter), 
\end{align}
other choices of bounding $\qt$ can also be used in EM \cite{Neal:Jordan:98}.

\section{Bilinear Adaptive VAMP}

We now propose an algorithm that approximates the quantities in \eqref{ThetaML}-\eqref{Xmmse}, i.e., the ML estimate of $\Thetabf=\{\thetabf_A,\thetabf_x,\gamma_w\}$ and the MMSE estimate of $\Xb$ under the statistical model \eqref{model}.
We start by developing Bilinear EM-VAMP and then add ``variance auto-tuning'' to obtain Bilinear Adaptive VAMP (BAd-VAMP).

\subsection{Bilinear EM-VAMP}

From the descriptions of VAMP and EM in \secref{background}, we see that they both minimize the same Gibbs free energy cost $J(q_1,q_2,q_3;\Thetabf)$ from \eqref{J_cost}, but w.r.t.\ different variables; 
VAMP minimizes $J$ w.r.t.\ the moment-constrained beliefs $\{q_1,q_2,q_3\}$ for a given $\Thetabf$, while EM minimizes $J$ w.r.t.\ $\Thetabf$ for a given $\{q_1,q_2,q_3\}$. 
As a result, the two approaches can be straightforwardly merged for \emph{joint} estimation of $\{q_1,q_2,q_3\}$ and $\Thetabf$. 
In doing so, the goal is to solve the optimization problem
\begin{subequations} \label{eq:EMVAMPopt}
\begin{align}
&\arg\min_{\Thetabf,q_1}\min_{q_2}\max_{q_3} J(q_1,q_2,q_3;\Thetabf)\\
&\text{s.t.}\quad \Exp[\xbf|q_1] = \Exp[\xbf|q_2] = \Exp[\xbf|q_3]\\
&\tr\{\cov[\xbf|q_1]\} = \tr\{\cov[\xbf|q_2]\} = \tr\{\cov[\xbf|q_3]\} ,
\end{align}
\end{subequations}
and the proposed methodology is to ``interleave'' the VAMP and EM algorithms, as specified in \algref{bemvamp}.
There, the estimation functions $\gb_1$ in lines~\ref{line:x1_bem}-\ref{line:eta1_bem} and $\gb_{2,l}$ in lines \ref{line:x2_bem}-\ref{line:eta2_bem} are defined as
\begin{align}
\lefteqn{\gb_1(\rb_{1,l},\gamma_{1,l};\thetabf_x)}\nonumber\\
&\defn \frac{\int \xb\,\px(\xb;\thetabf_x)\normal(\xb;\rb_{1,l},\Ib/\gamma_{1,l})\dif\xb}{\int \px(\xb;\thetabf_x)\normal(\xb;\rb_{1,l},\Ib/\gamma_{1,l})\dif\xb}
\label{eq:g1l_defn} \\
\lefteqn{\gb_{2,l}(\rb_{2,l},\gamma_{2,l};\thetabf_A,\gamma_w)}\nonumber\\
&\defn \frac{\int \xb\,\pygx(\yb_l|\xb;\thetabf_A,\gamma_w)\normal(\xb;\rb_{2,l},\Ib/\gamma_{2,l})\dif\xb}{\int \pygx(\yb_l|\xb;\thetabf_A,\gamma_w)\normal(\xb;\rb_{2,l},\Ib/\gamma_{2,l})\dif\xb}
\label{eq:g2l_defn} ,
\end{align}
The other lines in \algref{bemvamp} will be detailed in \secref{details}.

\begin{algorithm}[t]
\caption{Bilinear EM-VAMP}
\label{alg:bemvamp}
\begin{algorithmic}[1]
\Initialize {$\forall l: \rb_{1,l}^0,\gamma_{1,l}^0, 
              \thetabf_x^0, \thetabf_A^{0}, \gamma_w^0$}
\For{$t=0,\dots,T_{\max}$}
\State $\forall l:{\xb}_{1,l}^\iter=\gb_1(\rb_{1,l}^\iter,\gamma_{1,l}^\iter;\thetabf_x^\iter)$ \label{line:x1_bem}
\State $\forall l:1/{\eta}_{1,l}^\iter= \langle\gb_1'(\rb_{1,l}^\iter,\gamma_{1,l}^\iter;\thetabf_x^\iter)\rangle/\gamma_{1,l}^\iter$ \label{line:eta1_bem}
\State $q_1^\iter(\Xbf) \propto \prod_{l=1}^L \px(\xbf_l;\thetabf_x^\iter) e^{-\frac{1}{2}\gamma_{1,l}^\iter\|\xbf_l-\rbf_{1,l}^\iter\|^2}$ \label{line:q1_bem}
\State $\thetabf_x^\iters = \argmax_{\thetabf_x} \Exp[\ln \pX(\Xbf;\thetabf_x)|q_1^\iter]$ \label{line:thetax_bem}
\State $\forall l:{\gamma_{2,l}^{\iter}=\eta_{1,l}^{\iter}-\gamma_{1,l}^{\iter}}$
\State ${\forall l:\rb_{2,l}^\iter=(\eta_{1,l}^\iter\xb_{1,l}^\iter-\gamma_{1,l}^\iter\rb_{1,l}^\iter)/\gamma_{2,l}^\iter}$
\State $\forall l:{\xb}_{2,l}^\iter=\gb_{2,l}(\rb_{2,l}^\iter,\gamma_{2,l}^\iter;\thetabf_A^\iter,\gamma_w^\iter)$ \label{line:x2_bem}
\State $\forall l:1/{\eta}_{2,l}^\iter= \langle\gb_{2,l}'(\rb_{2,l}^\iter,\gamma_{2,l}^\iter;\thetabf_A^\iter,\gamma_w^\iter)\rangle/\gamma_{2,l}^\iter$ \label{line:eta2_bem}
\State $q_2^\iter(\Xbf) \propto \prod_{l=1}^L \pygx(\ybf_l|\xbf_l;\thetabf_A^\iter,\gamma_w^\iter) e^{-\frac{1}{2}\gamma_{2,l}^\iter\|\xbf_l-\rbf_{2,l}^\iter\|^2}$ \label{line:q2_bem}
\State $\thetabf_A^\iters = \argmax_{\thetabf_A} \Exp[\ln \pYgX(\Ybf|\Xbf;\thetabf_A,\gamma_w^\iter)|\Ybf,q_2^\iter]$ \label{line:thetaA_bem}
\State $\gamma_w^\iters = \argmax_{\gamma_w} \Exp[\ln \pYgX(\Ybf|\Xbf;\thetabf_A^\iters,\gamma_w)|\Ybf,q_2^\iter]$ \label{line:gammaw_bem}
\State $\forall l: \gamma_{1,l}^{\iters}=\eta_{2,l}^{\iter}-\gamma_{2,l}^{\iter}$
\State $\forall l: {\rb_{1,l}^\iters=(\eta_{2,l}^\iter\xb_{2,l}^\iter-\gamma_{2,l}^\iter\rb_{2,l}^\iter) /\gamma_{1,l}^\iters}$
\EndFor 
\end{algorithmic}
\end{algorithm}

\subsection{Bilinear Adaptive VAMP}

The VAMP state-evolution from \cite[Eq. (34), (35)]{Rangan:VAMP} shows that when 
i) $\Abf(\thetabf_A^\iter)$ is \textb{infinitely} large and right-rotationally invariant and 
ii) the estimation functions $\gbf_1$ and $\gbf_{2,l}$ are ``matched'' (i.e., MMSE) for the statistical model generating $(\Xbf,\Ybf)$,
the VAMP quantities $\{(\rbf_{i,l}^\iter,\gamma_{i,l}^\iter)\}_{l=1}^L$ for $i=1,2$ obey
\textb{\begin{subequations}
\begin{align}
\rbf_{1,l}^\iter
&= \xbf_l + \normal(\zero,\Ib/\gamma_{1,l}^\iter) ~\forall l\\
\xbf_{l}
&= \rbf_{2,l}^\iter + \normal(\zero,\Ib/\gamma_{2,l}^\iter) ~\forall l
\label{eq:r1},
\end{align}
\end{subequations}}
where $\xbf_l$ is the $l$th column of the true signal realization $\Xb$ that we aim to recover.
That is, $\rbf_{1,l}^\iter$ is an AWGN-corrupted version of the true signal $\xbf_l$ with \emph{known} AWGN precision $\gamma_{1,l}^\iter$,
and \textb{the true signal $\xbf_l$ is an AWGN-corrupted version of $\rbf_{2,l}^\iter$ with \emph{known} AWGN precision $\gamma_{2,l}^\iter$}. 
In the context of EM-VAMP under \eqref{model}, this ``matched'' condition requires that $\thetabf_A^\iter$, $\thetabf_x^\iter$, and $\gamma_w^\iter$ are all perfect estimates.
When $\thetabf_A^\iter$, $\thetabf_x^\iter$, or $\gamma_w^\iter$ are \emph{not} perfe1t, so that $\gbf_1$ and $\gbf_{2,l}$ are mismatched, the VAMP state-evolution shows that $\rbf_{1,l}^\iter$ is still an AWGN corrupted version of $\xbf_l$, but with an AWGN precision \emph{different} than $\gamma_{1,l}^\iter$.
The impact on EM-VAMP is the following.
While the algorithm is trying to learn $\Thetabf=\{\thetabf_A, \thetabf_x, \gamma_w\}$, the value of $\gamma_{i,l}^\iter$ does \emph{not} correctly characterize the noise precision in $\rbf_{i,l}^\iter$.
As a result, the beliefs $q_1^\iter$ and $q_2^\iter$ in lines~\ref{line:q1_bem} and \ref{line:q2_bem} of \algref{bemvamp} become mismatched, which compromises the EM updates of $\Thetabf^\iter$.

To remedy this situation, it was proposed in \cite{Kamilov:TIT:14} (in the context of EM-GAMP \cite{Vila:TSP:13}) to explicitly estimate the precision of the AWGN corruption on $\rbf_{1,l}^\iter$ and $\rbf_{2,l}^\iter$ and use it in place of the AMP-supplied estimates $\gamma_{1,l}^\iter$ and $\gamma_{2,l}^\iter$.
This approach was coined ``Adaptive'' GAMP in \cite{Kamilov:TIT:14}
and later extended to (linear) Adaptive VAMP in \cite{Fletcher:NIPS:17}.

For Bilinear Adaptive VAMP, the first goal is to replace the estimation of $\thetabf_x^\iter$ in \lineref{thetax_bem} of \algref{bemvamp} with the joint ML estimation
\begin{align}
(\thetabf_x^\iter,\gammabf_1^\iter) 
&= \arg\max_{\thetabf_x,\gammabf_1} 
p(\Rb_1^\iter;\gammabf_1,\thetabf_x)
\label{eq:ML_auto-tuning}
\end{align}
under the statistical model 
\begin{align}
\rb_{1,l}^\iter
&= \xbf_l + \normal(\zero,\Ib/\gamma_{1,l})~\forall l, 
& \quad\xbf_l \sim \px(\cdot;\thetabf_x)~\forall l,
\end{align}
with independence across $l=1,\dots,L$.
For this subproblem, we propose to use (inner) EM iterations indexed by $\tau$, i.e.,
\begin{align}
\lefteqn{ (\thetabf^{\taup1}_x,\gammabf^{\taup1}_1) }\nonumber\\
&= \arg\max_{\thetabf_x,\gammabf_1} \Exp\big[\ln p(\Xbf,\Rb_1^\iter;\gammabf_1,\thetabf_x) \,\big|\, \Rb_1^\iter; \gammabf_1^\tau,\thetabf_x^\tau \big]
\label{eq:EM_auto-tuning}\\
&= \arg\max_{\thetabf_x,\gammabf_1} \Big\{
\Exp\big[\ln p(\Xbf;\thetabf_x) \,\big|\, \Rb_1^\iter; \gammabf_1^\tau,\thetabf_x^\tau \big] 
\nonumber\\&\quad
+ \Exp\big[\ln p(\Rb_1^\iter|\Xb;\gammabf_1) \,\big|\, \Rb_1^\iter; \gammabf_1^\tau,\thetabf_x^\tau \big] 
\Big\} .
\end{align} 
The previous optimization decouples into
\begin{align} 
\thetabf^{\taup1}_x
&= \arg\max_{\thetabf_x} 
\Exp\big[\ln p(\Xbf;\thetabf_x) \,\big|\, \Rb_1^\iter; \gammabf_1^\tau,\thetabf_x^\tau \big] 
\end{align} 
and
\begin{align} 
\lefteqn{ \gammabf^{\taup1}_1
= \arg\max_{\gammabf_1} 
\Exp\big[\ln p(\Rb_1^\iter|\Xb;\gammabf_1) \,\big|\, \Rb_1^\iter; \gammabf_1^\tau,\thetabf_x^\tau \big] }\\
&= \arg\max_{\gammabf_1} \sum_{l=1}^L 
\Exp\big[\ln p(\rb_{1,l}^\iter|\xb_l;\gamma_{1,l}) \,\big|\, \rb_{1,l}^\iter; \gamma_{1,l}^\tau,\thetabf_x^\tau \big] ,
\end{align} 
where the latter optimization decouples further into 
\begin{align} 
\gamma^{\taup1}_{1,l}
&= \arg\max_{\gamma_{1,l}}
\bigg\{ \frac{N}{2}\ln\gamma_{1,l}
\nonumber\\&\qquad
- \frac{\gamma_{1,l}}{2}\Exp\big[\|\xbf_l-\rb_{1,l}^\iter\|_2^2 \,\big|\, \rb_{1,l}^\iter; \gamma_{1,l}^\tau,\thetabf_x^\tau\big]
\bigg\} \\
&= N\left\{\Exp\big[\|\xbf_l-\rb_{1,l}^\iter\|_2^2 \,\big|\, \rb_{1,l}^\iter; \gamma_{1,l}^\tau,\thetabf_x^\tau\big]\right\}^{-1} \\
&= \left\{\frac{1}{N}\sum_{n=1}^N \Exp\big[(x_{nl}-r_{1,nl}^\iter)^2 \,\big|\, \rb_{1,l}^\iter; \gamma_{1,l}^\tau,\thetabf_x^\tau\big]\right\}^{-1} \\
&= \left\{ \frac{1}{N} \|\xb_{1,l}^{\tau}-\rb_{1,l}^\iter\|^2 + \frac{1}{\eta_{1,l}^{\tau}} \right\}^{-1},
\end{align} 
for $l=1,\dots,L$ and
\begin{align} 
\xb_{1,l}^{\tau}
&\defn \Exp\big[\xb_l \big| \rb_{1,l}^\iter; \gamma_{1,l}^\tau,\thetabf_x^\tau \big] \\
&= \gb_1(\rb_{1,l}^\iter,\gamma_{1,l}^\tau;\thetabf_x^\tau)\\
1/\eta_{1,l}^{\tau}
&\defn \tr\big\{\cov\big[\xb_l \big| \rb_{1,l}^\iter; \gamma_{1,l}^\tau,\thetabf_x^\tau \big] \big\}/N \\
&= \langle \gb_1'(\rb_{1,l}^\iter,\gamma_{1,l}^\tau;\thetabf_x^\tau) \rangle/\gamma_{1,l}^\tau .
\end{align} 

Above, we detailed the re-estimation of $\gammabf_1^\iter$.
A similar procedure can be used for re-estimation of $\gammabf_2^\iter$.
The resulting Bilinear Adaptive VAMP (BAd-VAMP) is summarized in \algref{badvamp} using $\tau_{1,\max}$ EM iterations for the first inner loop and $\tau_{2,\max}$ EM iterations for the second inner loop.
To avoid the complications of a dual-index notation (i.e., $t$ and $\tau$), we use only the single index $t$ in \algref{badvamp} and over-write the quantities in each inner loop.
Note that, when $\tau_{1,\max}=\tau_{2,\max}=0$, BAd-VAMP (i.e., \algref{badvamp}) reduces to bilinear EM-VAMP (i.e., \algref{bemvamp}).

\begin{algorithm}[t]
\caption{Bilinear Adaptive VAMP}
\label{alg:badvamp}
\begin{algorithmic}[1]
\Initialize {$\forall l: \rb_{1,l}^0,\gamma_{1,l}^0, 
              \thetabf_x^0, \thetabf_A^{0}, \gamma_w^0$}
\For{$t=0,\dots,T_{\max}$}
\For{$\tau=0,\dots,\tau_{1,\max}$}
\State $\forall l:{\xb}_{1,l}^\iter \leftarrow \gb_1(\rb_{1,l}^\iter,\gamma_{1,l}^\iter;\thetabf_x^\iter)$ 
\State $\forall l:1/{\eta}_{1,l}^\iter \leftarrow \langle\gb_1'(\rb_{1,l}^\iter,\gamma_{1,l}^\iter;\thetabf_x^\iter)\rangle/\gamma_{1,l}^\iter$ 
\State $\forall l:1/\gamma_{1,l}^\iter \leftarrow \frac{1}{N} \|\xb_{1,l}^\iter-\rb_{1,l}^\iter\|^2 + 1/\eta_{1,l}^\iter$
\State $q_1^\iter(\Xbf) \propto \prod_{l=1}^L \px(\xbf_l;\thetabf_x^\iter) e^{-\frac{1}{2}\gamma_{1,l}^\iter\|\xbf_l-\rbf_{1,l}^\iter\|^2}$ 
\State $\thetabf_x^\iter \leftarrow \argmax_{\thetabf_x} \Exp[\ln \pX(\Xbf;\thetabf_x)|q_1^\iter]$ 
\EndFor 
\State $\thetabf_x^\iters = \thetabf_x^\iter$
\State $\forall l:{\gamma_{2,l}^{\iter}=\eta_{1,l}^{\iter}-\gamma_{1,l}^{\iter}}$ \label{line:gamma2_bad}
\State ${\forall l:\rb_{2,l}^\iter=(\eta_{1,l}^\iter\xb_{1,l}^\iter-\gamma_{1,l}^\iter\rb_{1,l}^\iter)/\gamma_{2,l}^\iter}$ 
\For{$\tau=0,\dots,\tau_{2,\max}$}
\State $\forall l:{\xb}_{2,l}^\iter \leftarrow \gb_{2,l}(\rb_{2,l}^\iter,\gamma_{2,l}^\iter;\thetabf_A^\iter,\gamma_w^\iter)$ 
\State $\forall l:1/{\eta}_{2,l}^\iter \leftarrow \langle\gb_{2,l}'(\rb_{2,l}^\iter,\gamma_{2,l}^\iter;\thetabf_A^\iter,\gamma_w^\iter)\rangle/\gamma_{2,l}^\iter$ 
\State $\forall l:1/\gamma_{2,l}^\iter \leftarrow \frac{1}{N} \|\xb_{2,l}^\iter-\rb_{2,l}^\iter\|^2 + 1/\eta_{2,l}^\iter$
\State $q_2^\iter(\Xbf) \propto \prod_{l} \pygx(\ybf_l|\xbf_l;\thetabf_A^\iter,\gamma_w^\iter) e^{-\frac{1}{2}\gamma_{2,l}^\iter\|\xbf_l-\rbf_{2,l}^\iter\|^2}$ 
\State $\thetabf_A^\iter \leftarrow \argmax_{\thetabf_A} \Exp[\ln \pYgX(\Ybf|\Xbf;\thetabf_A,\gamma_w^\iter)|\Ybf,q_2^\iter]$\label{line:thetaA_bad}%
\State $\gamma_w^\iter \leftarrow \argmax_{\gamma_w} \Exp[\ln \pYgX(\Ybf|\Xbf;\thetabf_A^\iter,\gamma_w)|\Ybf,q_2^\iter]$\label{line:gammaw_bad}%
\EndFor 
\State $\thetabf_A^\iters = \thetabf_A^\iter$
\State $\gamma_w^\iters = \gamma_w^\iter$
\State $\forall l: \gamma_{1,l}^{\iters}=\eta_{2,l}^{\iter}-\gamma_{2,l}^{\iter}$ \label{line:gamma1_bad}
\State $\forall l: {\rb_{1,l}^\iters=(\eta_{2,l}^\iter\xb_{2,l}^\iter-\gamma_{2,l}^\iter\rb_{2,l}^\iter) /\gamma_{1,l}^\iters}$ \label{line:r1_bad}
\EndFor 
\end{algorithmic}
\end{algorithm}

\subsection{Algorithm Details} \label{sec:details}

We now provide additional details on the steps in \algref{badvamp}.

\subsubsection{Estimating \texorpdfstring{$\Xbf$}{X}}

Recalling the definition of $\gb_{2,l}(\cdot)$ in \eqref{g2l_defn},
the form of $\pygx$ in \eqref{py|x} implies that
\begin{eqnarray}
\gb_{2,l}(\rb_{2,l}^\iter,\gamma_{2,l}^\iter;\thetabf_A^\iter,\gamma_w^\iter)
&=& \Cbf_l^\iter
\big(\gamma_{2,l}^\iter\rb_{2,l}^\iter+\gamma_w^\iter\Ab(\thetabf_A^\iter)\tran\yb_l\big) \qquad \\
\langle \gb'_{2,l}(\rb_{2,l}^\iter,\gamma_{2,l}^\iter;\thetabf_A^\iter,\gamma_w^\iter) \rangle
&=& \gamma_{2,l}^\iter \tr\big\{ \Cbf_l^\iter \big\}/N
\end{eqnarray}
for 
\begin{align}
\Cbf_l^\iter
&\defn \big(\gamma_{2,l}^\iter\Ib_N + \gamma_w^\iter\Ab(\thetabf_A^\iter)\tran\Ab(\thetabf_A^\iter)\big)^{-1} 
\label{eq:Ql} .
\end{align}

To avoid computing a separate matrix inverse \eqref{Ql} for each $l=1,\dots,L$, one could instead compute the eigenvalue decomposition
\begin{align}
\Ab(\thetabf_A^\iter)\tran\Ab(\thetabf_A^\iter)
=\Ub^\iter\Diag(\sb^\iter){\Ub^\iter}\tran ,
\end{align}
and then leverage the fact that
\begin{align}
\Cbf_l^\iter
=\Ub^\iter\Diag(\gamma_{2,l}^\iter\one + \gamma_w^\iter\sb^\iter)^{-1}{\Ub^\iter}\tran ,
\end{align}
which reduces to the inversion of a diagonal matrix for each $l=1,\dots,L$.

\subsubsection{Learning \texorpdfstring{$\thetabf_A$}{theta-A}}

We now provide details on the update of $\thetabf_A$ and $\gamma_w$ in lines~\ref{line:thetaA_bad}-\ref{line:gammaw_bad} of \algref{badvamp}. 
Given the form of $\pYgX$ in \eqref{pY|X}-\eqref{py|x}, we have that
\begin{align}
\lefteqn{ \ln \pYgX(\Yb|\Xb;\thetabf_A,\gamma_w) }\nonumber\\
&= \tfrac{ML}{2}\ln \gamma_w
   -\tfrac{\gamma_w}{2} \|\Yb-\Ab(\thetabf_A)\Xbf\|_F^2 + \text{const} \\
&= \tfrac{ML}{2}\ln \gamma_w
   -\tfrac{\gamma_w}{2} \big( 
   \tr\{\Yb\Yb\tran\}
   -2\tr\{\Ab(\thetabf_A)\Xbf\Ybf\tran\} 
   \nonumber\\&\quad
   + \tr\{\Ab(\thetabf_A)\Xbf\Xbf\tran\Ab(\thetabf_A)\tran\} 
   \big) + \text{const} .
\end{align}
Since
\begin{align}
\Exp\big[\Xbf\big|q_2^\iter\big] 
&= \Xbf_2^\iter \\
\Exp\big[\Xbf\Xbf\tran\big|q_2^\iter\big] 
&= \sum_{l=1}^L \Exp\big[\xbf_l\xbf_l\tran \big| q_{2,l}^\iter\big] 
= \Xbf_2^\iter{\Xbf_2^\iter}\tran + \underbrace{\sum_{l=1}^L \Cbf_l^\iter}_{\displaystyle \defn \Cbf^t} ,
\end{align}
we have that
\begin{align}
\lefteqn{ \Exp\big[ \ln \pYgX(\Yb|\Xb;\thetabf_A,\gamma_w) \big| \Ybf,q_2^\iter\big] }\nonumber\\
&= \frac{ML}{2}\ln \gamma_w
   -\frac{\gamma_w}{2} \Big( 
   \tr\big\{\Yb\Yb\tran\big\}
   -2\tr\big\{\Ab(\thetabf_A)\Xbf_2^\iter\Ybf\tran\big\} 
   \nonumber\\&\quad
   + \tr\big\{\Ab(\thetabf_A)\Xbf_2^\iter{\Xbf_2^\iter}\tran\Ab(\thetabf_A)\tran\big\} 
   + \tr\big\{\Ab(\thetabf_A)\Cbf^\iter\Ab(\thetabf_A)\tran\big\} 
   \Big) 
\label{eq:ElnpYgXtr}\\
&= \frac{ML}{2}\ln \gamma_w
   -\frac{\gamma_w}{2} \Big( 
   \|\Yb-\Ab(\thetabf_A)\Xb_2^t\|_F^2
   \nonumber\\&\quad
   + \tr\big\{\Ab(\thetabf_A)\Cbf^\iter\Ab(\thetabf_A)\tran\big\} 
   \Big) 
   + \text{const} 
\label{eq:ElnpYgX}.
\end{align}

To maximize \eqref{ElnpYgX} over $\thetabf_A=[\theta_{A,1},\dots,\theta_{A,Q}]$ with fixed $\gamma_w$, we consider the affine-linear model 
\begin{align}
\Ab(\thetabf_A) &= \Ab_0 + \sum_{i=1}^{Q}\theta_{A,i}\Ab_i
\label{eq:A_affine} ,
\end{align}
noting that non-linear models could be handled using similar techniques.
Plugging \eqref{A_affine} into \eqref{ElnpYgXtr}, we get 
\begin{align}
\lefteqn{ \Exp\big[ \ln \pYgX(\Yb|\Xb;\thetabf_A,\gamma_w) \big| \Ybf,q_2^\iter\big] }\nonumber\\
&= \text{const} -
\frac{\gamma_w}{2}\sum_{i=1}^{Q}\sum_{j=1}^{Q}
\theta_{A,i}
\tr\{\Abf_i(\Cb^\iter+\Xb_2^\iter{\Xb_2^\iter}\tran)\Abf_j\tran\}
\theta_{A,j}
\nonumber\\&\quad
-\gamma_w\sum_{i=1}^{Q} \theta_{A,i}\big(
\tr\{\Ab_i(\Cb^\iter+\Xb_2^\iter{\Xb_2^\iter}\tran)\Ab_0\tran\}
-\tr\{\Ab_i\Xb_2^\iter\Yb\tran\}
\big) \\
&=-\frac{\gamma_w}{2}\Big(\thetabf_A\tran\Hb^\iter\thetabf_A
  -2\thetabf_A\tran\betabf^\iter \Big) + \text{const}
\label{eq:ElnpYgXtheta} 
\end{align}
for
\begin{eqnarray}
[\Hbf^\iter]_{ij}
&=&\tr\big\{\Ab_i(\Cb^\iter+\Xb_2^\iter{\Xb_2^\iter}\tran)\Ab_j\tran\big\} \\
&=&\tr\big\{\Ab_j\tran\Ab_i(\Cb^\iter+\Xb_2^\iter{\Xb_2^\iter}\tran)\big\}
\label{eq:H_defn}
\end{eqnarray}
and
\begin{eqnarray}
{[\betabf^\iter]}_i
&=& \tr\big\{\Ab_i\Xb_2^\iter\Yb\tran\big\}
-\tr\big\{\Ab_i(\Cbf^t + \Xbf_2^\iter{\Xbf_2^\iter}\tran)\Abf_0\tran\big\} \\
&=& \tr\big\{\Yb\tran\Ab_i\Xb_2^\iter\big\}
-\tr\big\{\Ab_0\tran\Ab_i(\Cbf^t + \Xbf_2^\iter{\Xbf_2^\iter}\tran)\big\}
\label{eq:beta_defn} , \quad
\end{eqnarray}
where $\Ab_j\tran\Ab_i$ and $\Yb\tran\Ab_i$ can be pre-computed.
Zeroing the gradient of \eqref{ElnpYgXtheta} w.r.t.\ $\thetabf_A$, we find that the maximizer is
\begin{align}
\thetabf_A^\iters
&=(\Hb^\iter)^{-1}\betabf^\iter
\label{eq:ga_affine} .
\end{align}

A special case of \eqref{A_affine} is where $\Ab(\cdot)$ has no structure, i.e.,
\begin{align}
\Ab(\thetabf_A) 
&= \sum_{m=1}^M\sum_{n=1}^N \theta_{A,m,n}\eb_m\eb_n\tran
\label{eq:A_unstructured} .
\end{align}
where $\eb_m$ denotes the $m$th standard basis vector. 
In this case, it can be shown that
\begin{align}
\Ab(\thetabf_A^\iters) 
&= \Yb{\Xb_2^\iter}\tran\big(\Cb^\iter+\Xb_2^\iter{\Xb_2^\iter}\tran\big)^{-1}.
\end{align}

\subsubsection{Learning \texorpdfstring{$\gamma_w$}{gamma-w}}

To maximize \eqref{ElnpYgX} over $\gamma_w$ with fixed $\thetabf_A=\thetabf_A^\iters$, we search for the values of $\gamma_w$ that zero the derivative of \eqref{ElnpYgX}.
The unique solution is straightforwardly shown to be
\begin{align}
1/\gamma_w^\iters
&= \frac{1}{ML} \Big( \|\Yb-\Ab(\thetabf_A^\iters)\Xb_2^t\|_F^2
   \nonumber\\&\quad
   + \tr\big\{\Ab(\thetabf_A^\iters)\Cbf^\iter\Ab(\thetabf_A^\iters)\tran\big\} 
   \Big) .
\end{align}

\subsubsection{Summary}
In \algref{badvamp_detailed}, BAd-VAMP is rewritten with detailed expressions for the updates of $\xb_{2,l}^\iter$, $\eta_{2,l}^\iter$, $\thetabf_A^\iter$, and $\gamma_w^\iter$.

\begin{algorithm}[t]
\caption{Bilinear Adaptive VAMP (Detailed)}
\label{alg:badvamp_detailed}
\begin{algorithmic}[1]
\Initialize {$\forall l: \rb_{1,l}^0,\gamma_{1,l}^0, 
              \thetabf_x^0, \thetabf_A^{0}, \gamma_w^0$}
\For{$t=0,\dots,T_{\max}$}
\For{$\tau=0,\dots,\tau_{1,\max}$}
\State $\forall l:{\xb}_{1,l}^\iter \leftarrow \gb_1(\rb_{1,l}^\iter,\gamma_{1,l}^\iter;\thetabf_x^\iter)$ 
\State $\forall l:1/{\eta}_{1,l}^\iter \leftarrow \langle\gb_1'(\rb_{1,l}^\iter,\gamma_{1,l}^\iter;\thetabf_x^\iter)\rangle/\gamma_{1,l}^\iter$ 
\State $\forall l:1/\gamma_{1,l}^\iter \leftarrow \frac{1}{N} \|\xb_{1,l}^\iter-\rb_{1,l}^\iter\|^2 + 1/\eta_{1,l}^\iter$
\State $q_1^\iter(\Xbf) \propto \prod_{l=1}^L \px(\xbf_l;\thetabf_x^\iter) e^{-\frac{1}{2}\gamma_{1,l}^\iter\|\xbf_l-\rbf_{1,l}^\iter\|^2}$ 
\State $\thetabf_x^\iter \leftarrow \argmax_{\thetabf_x} \Exp[\ln \pX(\Xbf;\thetabf_x)|q_1^\iter]$ 
\EndFor 
\State $\thetabf_x^\iters = \thetabf_x^\iter$
\State $\forall l:{\gamma_{2,l}^{\iter}=\eta_{1,l}^{\iter}-\gamma_{1,l}^{\iter}}$ \label{line:gamma2}
\State ${\forall l:\rb_{2,l}^\iter=(\eta_{1,l}^\iter\xb_{1,l}^\iter-\gamma_{1,l}^\iter\rb_{1,l}^\iter)/\gamma_{2,l}^\iter}$ 
\For{$\tau=0,\dots,\tau_{2,\max}$}
\State $\forall l: \Cb_{l}^\iter \leftarrow \big(\gamma_{2,l}^\iter\Ib_N+\gamma_w^\iter\Ab(\thetabf_A^\iter)\tran\Ab(\thetabf_A^\iter)\big)^{-1}$ 
\State $\forall l:\xb_{2,l}^\iter \leftarrow \Cb_l^\iter\big( \gamma_{2,l}^\iter\rb_{2,l}^\iter +\gamma_w^\iter\Ab(\thetabf_A^\iter)\tran\yb_l \big)$ 
\State $\forall l: 1/\eta_{2,l}^\iter \leftarrow \tr\{\Cb_l^\iter\}/N$ 
\State $\Cb^\iter \leftarrow \sum_{l=1}^L\Cb_l^\iter$ 
\State $\forall i,j: [\Hbf^\iter]_{ij} \leftarrow \tr\big\{\Ab_j\tran\Ab_i(\Cb^\iter+\Xb_2^\iter{\Xb_2^\iter}\tran)\big\}$ 
\State $\begin{array}{ll}
        \forall i:[\betabf^\iter]_i \leftarrow &\tr\big\{\Yb\tran\Ab_i\Xb_2^\iter\big\} \\
        &-\tr\big\{\Abf_0\tran\Ab_i(\Cbf^t + \Xbf_2^\iter{\Xbf_2^\iter}\tran)\big\}
        \end{array}$
\State $\thetabf_A^\iter \leftarrow (\Hb^\iter)^{-1}\betabf^\iter$
\State $\begin{array}{l@{}l}
        1/\gamma_w^\iter \leftarrow \frac{1}{ML}\big(&\|\Yb-\Ab(\thetabf^\iter)\Xb_2^\iter\|_F^2\\
        & +\tr\big\{\Ab(\thetabf_A^\iter)\Cb^\iter\Ab(\thetabf_A^\iter)\tran\big\}\big)
        \end{array}$ 
\EndFor 
\State $\thetabf_A^\iters = \thetabf_A^\iter$
\State $\gamma_w^\iters = \gamma_w^\iter$
\State $\forall l: \gamma_{1,l}^{\iters}=\eta_{2,l}^{\iter}-\gamma_{2,l}^{\iter}$ \label{line:gamma1}
\State $\forall l: \rb_{1,l}^\iters=(\eta_{2,l}^\iter\xb_{2,l}^\iter-\gamma_{2,l}^\iter\rb_{2,l}^\iter) / \gamma_{1,l}^\iters$ \label{line:r1}
\EndFor 
\end{algorithmic}
\end{algorithm}

\subsection{Algorithm Enhancements}

We now propose several enhancements to the BAd-VAMP algorithm presented in \algref{badvamp} and detailed in \algref{badvamp_detailed}. 

\subsubsection{Damping}
For fixed $\Thetabf^\iter$ and \textb{infinitely} large right-rotationally invariant $\Ab(\thetabf_A^\iter)$, the state-evolution of VAMP guarantees its convergence.
But when $\Ab(\thetabf_A^\iter)$ deviates from this assumption, damping the VAMP iterations can help maintain convergence \cite{Rangan:VAMP}. 
With damping, lines~\ref{line:gamma1}-\ref{line:r1} of \algref{badvamp_detailed} (or lines~\ref{line:gamma1_bad}-\ref{line:r1_bad} of \algref{badvamp}) would be replaced by
\begin{align}
\gamma_{1,l}^{\iters}
&= (1-\zeta)\gamma_{1,l}^\iter + \zeta(\eta_{2,l}^{\iter}-\gamma_{2,l}^{\iter}) \\
\rb_{1,l}^{\iters}
&=(1-\zeta)\rb_{1,l}^\iter + \zeta(\eta_{2,l}^{\iter}\xb_{2,l}^{\iter}-\gamma_{2,l}^\iter\rb_{2,l}^\iter) /(\eta_{2,l}^{\iter}-\gamma_{2,l}^{\iter})
\end{align}
for some $\zeta\in(0,1)$.
The case $\zeta=1$ corresponds to no damping.

\subsubsection{Negative precisions}
Sometimes the precisions $\{\gamma_{1,l},\gamma_{2,l}\}_l$ can be negative. We suggest to restrict the precisions to the interval $[\gamma_{\min},\infty)$, for very small $\gamma_{\min}>0$, in lines~\ref{line:gamma2} and~\ref{line:gamma1} of \algref{badvamp_detailed} (or lines~\ref{line:gamma2_bad} and~\ref{line:gamma1_bad} of \algref{badvamp}).

\subsubsection{Restarts}
Due to the non-convex nature of the bilinear inference problem, the algorithm may get stuck at local minima or slowed by saddle points. 
To mitigate these issues, it sometimes helps to restart the algorithm. 
For each restart, we suggest to initialize $\thetabf_A^0$ at the final estimate of $\thetabf_A$ returned by the previous run. 

\subsection{Relation to Previous Work}

The proposed Bilinear Adaptive VAMP algorithm extends the (linear) Adaptive VAMP algorithm of \cite{Fletcher:NIPS:17} from the case where $\Abf(\thetabf_A)$ is known to the case where $\Abf(\thetabf_A)$ is unknown.
In the known-$\Abf(\thetabf_A)$ setting, where $\Abf(\thetabf_A)$ is \textb{infinitely} large and right-rotationally invariant, it was rigorously established in \cite{Fletcher:NIPS:17} that Adaptive VAMP obeys a state-evolution similar to that of VAMP, and that its estimates of $\{\thetabf_x,\gamma_w\}$ are asymptotically consistent under certain identifiability conditions, i.e., they converge to the true values as $t\rightarrow\infty$.
As future work, it would be interesting to understand whether Bilinear Adaptive VAMP also obeys a state evolution for certain classes of $\Abf(\cdot)$.

The proposed BAd-VAMP algorithm targets the same class\footnote{Note that the BiGAMP algorithm \cite{Parker:TSP:14a}-\cite{Parker:TSP:14b} is a special case of the PBiGAMP algorithm \cite{Parker:JSTSP:16}.  BiGAMP applies to the recovery of $\Abf$ and $\Xbf$ from measurements of the form $\Ybf=\Abf\Xbf+\Wbf$, whereas PBiGAMP applies to the recovery of $\bbf$ and $\Xbf$ from $\Ybf=\Abf(\bbf)\Xbf+\Wbf$ under known affine linear $\Abf(\cdot)$.  Both can be combined with EM for hyperparameter learning.} of bilinear recovery problems as the EM-PBiGAMP algorithm from \cite{Parker:JSTSP:16}, and both leverage EM for automated hyperparameter tuning.
However, the ``AMP'' aspects of these algorithms are fundamentally different.
PBiGAMP treats the vectors $\{\abf_{i,j}\}$ in \eqref{bilinear} as i.i.d.\ Gaussian for its derivation, whereas BAd-VAMP treats the matrix $\Ab(\bbf)=\sum_{i=1}^Q b_i\Ab_i$ as right rotationally-invariant for its derivation.
The latter allows more freedom in the singular values of $\Ab(\bbf)$, which leads to increased robustness in practice, as demonstrated by the numerical experiments in \secref{simulations}.

BAd-VAMP and lifted VAMP both leverage the VAMP approach from \cite{Rangan:VAMP} to solve bilinear inference problems.
However, they do so in very different ways.
As discussed in \secref{intro}, lifted VAMP ``lifts'' the bilinear problem into a higher-dimensional linear problem, and then uses non-separable denoising to jointly estimate $\bbf$ and $\cbf$ in \eqref{linear}.  
An unfortunate consequence of lifting is a possibly significant increase in computational complexity and memory.
In contrast, BAd-VAMP avoids lifting, and it employs EM to estimate $\bbf$ and VAMP to estimate $\cbf$.
Interestingly, \secref{simulations} shows BAd-VAMP performing equal or better to lifted VAMP in all experiments.

\section{Numerical Simulations} \label{sec:simulations}

In this section, we present the results of numerical simulations that study the behavior of the BAd-VAMP algorithm from \algref{badvamp_detailed}, in comparison to other state-of-the-art algorithms, on several bilinear recovery problems.
In all cases, we ran BAd-VAMP with $\tau_{1,\max}=1$ and $\tau_{2,\max}=0$ inner EM iterations \textb{and we assumed that the signal prior $\px$ is fully known (i.e., $\thetabf_x$ is known)}.
We nominally used a damping coefficient of $\zeta=0.8$ and minimum precision of $\gamma_{\min}=10^{-6}$. 
\textb{We initialized BAd-VAMP by $\gamma_w^0=0.1$, $\gamma_{1,l}^0=10^{-3}~\forall l$, and we set $\rbf_{1,l}^0$ and $\thetabf_A^0$ to random vectors drawn i.i.d.\ from $\normal(0,10)$ and $\normal(0,1)$ respectively.
Unless otherwise noted, no restarts were used.} 

\subsection{CS with Matrix Uncertainty}
\label{sec:csmu_experiment}

In compressive sensing (CS) with matrix uncertainty \cite{Zhu:TSP:11}, 
the goal is to recover the $K$-sparse signal $\cb\in\Real^{N}$ and the uncertainty parameters $\bbf$ from measurements $\yb=\Ab(\bbf)\cb + \wb$ with $\wb\sim\normal(\zero,\Ib/\gamma_w)$. 
Here, $\Ab(\bbf)=\Ab_0 + \sum_{i=1}^{Q}b_i\Ab_i$, where $\{\Abf_i\}_{i=0}^{Q}$ are known.
For our experiments, we used $Q=10$, $N=256$, and $K=10$, and we selected $\gamma_w$ so that $\text{SNR}\defn \Exp[\|\Ab\cb\|^2]/\Exp[\|\wb\|^2]=40$ dB. 
Also, the uncertainty parameters $\bbf$ were drawn $\normal(\zero,\Ib)$, 
and $\cb$ was drawn with uniformly random support and with $K$ non-zero elements from $\normal(\zero,\Ibf)$. 
We measured performance using $\text{NMSE}(\hat{\bbf})\defn\|\hat{\bbf}-\bbf\|^2/\|\bbf\|^2$ and $\text{NMSE}(\hat{\cb})\defn\|\hat{\cb}-\cb\|^2/\|\cb\|^2$. 
As a reference, we considered two oracle estimators: the MMSE estimator for $\bbf$ assuming $\cb$ is known, and the MMSE estimator for $\cb$ assuming $\bbf$ and the support of $\cb$ are known.

For our first experiment, 
the elements of $\{\Abf_i\}_{i=1}^{Q}$ were drawn i.i.d.\ $\normal(0,1)$  
and the elements of $\Ab_0$ were drawn $\normal(0,20)$. 
\textb{BAd-VAMP was run for a maximum of $200$ iterations with a maximum of $2$ restarts and damping $\zeta=0.86$.}

\begin{figure}[t]
\centering
\newcommand{\sz}{0.7}
\psfrag{x1}[t][t][\sz]{\sf sampling ratio $M/N$} 
\psfrag{x2}[t][t][\sz]{\sf sampling ratio $M/N$} 
\psfrag{y1}[b][B][\sz]{\sf NMSE($\hat{\bbf}$) [dB]} 
\psfrag{y2}[b][B][\sz]{\sf NMSE($\hat{\cb}$) [dB]}
\psfrag{EM-VAMP}[l][l][0.43]{\sf \hspace{-0.4mm}BAd-VAMP}
\psfrag{P-BiG-AMP}[l][l][0.43]{\sf \hspace{-0.4mm}PBiGAMP}
\includegraphics[width=\figsize,trim=15mm 3mm 15mm 8mm,clip]{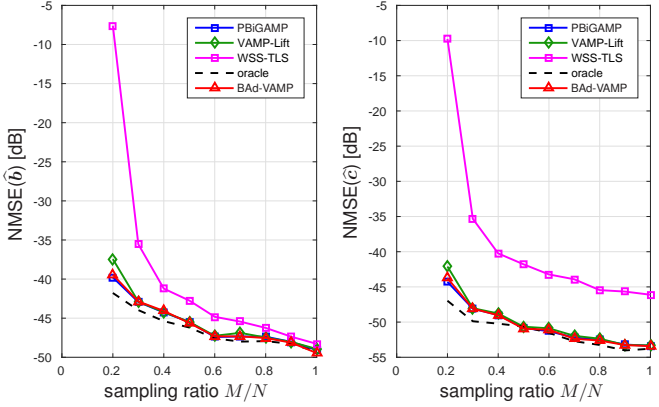}
\caption{CS with matrix uncertainty: Median NMSE (over 50 trials) on signal $\cb$ and uncertainty parameters $\bbf$ versus sampling ratio $M/N$.}
\label{fig:csmu_iid}
\end{figure}

\Figref{csmu_iid} shows that the AMP-based algorithms gave near-oracle performance for the tested range of $M/N$, although lifted VAMP performed slightly worse than the others when $M/N=0.2$.
In contrast, the performance of WSS-TLS from the award-winning paper \cite{Zhu:TSP:11} was significantly worse than the AMP approaches.
WSS-TLS aims to solve the non-convex optimization problem
\begin{align}
(\hat{\bbf},\hat{\cb}) 
&= \arg\min_{\bbf,\cb} \left\| \left(\Ab_0 + \sum_{i=1}^Q b_i \Ab_i \right)\cb - \yb\right\|^2 
\nonumber\\&\quad
+ \|\bbf\|^2/\gamma_w + \lambda \|\cb\|_1
\label{eq:WSSTLS}
\end{align}
using alternating minimization.
For WSS-TLS, we used oracle knowledge of $\gamma_w$, oracle tuning of the regularization parameter $\lambda$, and code from the authors' website.

\begin{figure}[t]
\centering
\newcommand{\sz}{0.7}
\psfrag{x1}[t][t][\sz]{\sf mean $\mu$ of $\Abf_i$} 
\psfrag{x2}[t][t][\sz]{\sf mean $\mu$ of $\Abf_i$} 
\psfrag{y1}[b][B][\sz]{\sf NMSE($\hat{\bbf}$) [dB]} 
\psfrag{y2}[b][B][\sz]{\sf NMSE($\hat{\cbf}$) [dB]}
\psfrag{EM-VAMP}[l][l][0.43]{\sf \hspace{-0.4mm}BAd-VAMP}
\psfrag{P-BiG-AMP}[l][l][0.43]{\sf \hspace{-0.4mm}PBiGAMP}
\includegraphics[width=\figsize,trim=15mm 3mm 15mm 8mm,clip]{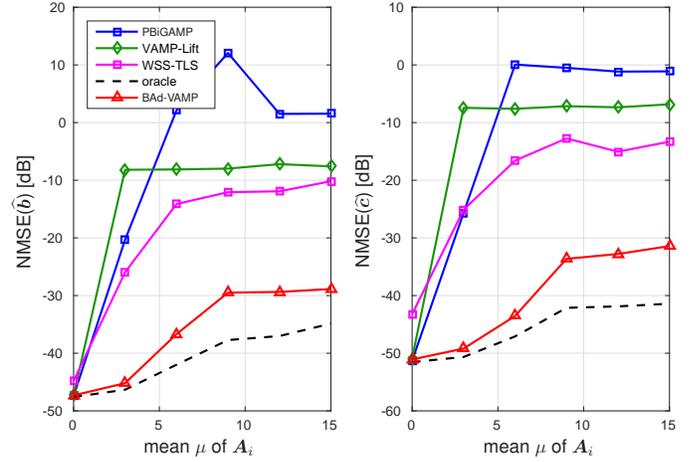}
\caption{CS with matrix uncertainty: Median NMSE (over 50 trials) on signal $\cb$ and uncertainty parameters $\bbf$ versus mean of matrices $\Abf_i$ at $M/N=0.6$.}  
\label{fig:csmu_Amean}
\end{figure}

For our second experiment, we tested algorithm robustness to non-zero mean in $\Abf(\bbf)$\footnote{\textb{For the simpler case where $\bbf$ is known and the objective is to recover $\cbf$ from $\ybf=\Abf\cbf + \wbf$, modifications of AMP that temporarily remove the mean from $\Abf$ have been proposed\cite{Vila:ICASSP:15}. However, it is not clear how to extend this approach to the bilinear problem of recovering $\bbf$ and $\cbf$ from $\ybf=\Abf(\bbf)\cbf+\wbf$. }}, since this is a known issue with many AMP algorithms \cite{Rangan:TIT:16a,Caltagirone:ISIT:14,Vila:ICASSP:15}. 
For this, we fixed the sampling ratio at $M/N=0.6$,
drew the elements of $\{\Abf_i\}_{i=1}^{Q}$ from i.i.d.\ $\normal(\mu,1)$, 
and drew the elements of $\Ab_0$ from i.i.d.\ $\normal(\mu,20)$. 
\Figref{csmu_Amean} reports the median NMSE versus mean $\mu$, and shows that BAd-VAMP is much more robust to $\mu>0$ than the other tested AMP algorithms as well as WSS-TLS.

\begin{figure}[t]
\centering
\newcommand{\sz}{0.7}
\psfrag{x1}[t][t][\sz]{\sf sampling ratio $M/N$} 
\psfrag{y1}[b][B][\sz]{\sf run-time [sec]} 
\psfrag{P-BiG-AMP}[l][l][0.5]{\sf \hspace{-0.25mm}PBiGAMP}
\psfrag{BAd-VAMP}[l][l][0.5]{\sf \hspace{-0.25mm}BAd-VAMP}
\psfrag{VAMP-Lift}[l][l][0.5]{\sf \hspace{-0.25mm}VAMP-Lift}
\psfrag{WSS-TLS}[l][l][0.5]{\sf \hspace{-0.25mm}WSS-TLS}
\includegraphics[width=\figsize,trim=5mm 5mm 5mm 5mm,clip]{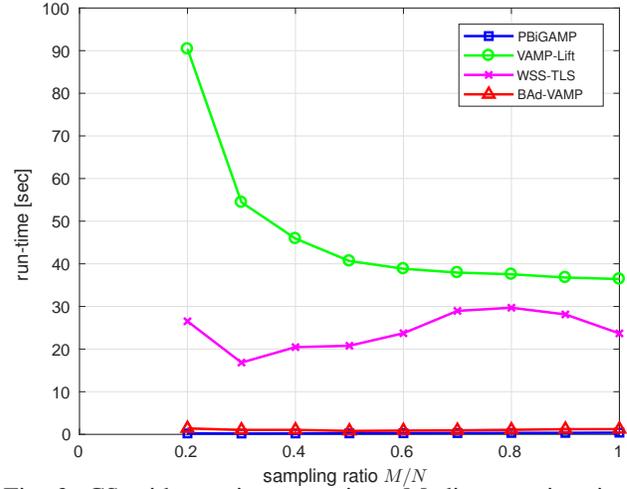}
\caption{CS with matrix uncertainty: Median run-time in seconds (over 10 trials) versus sampling ratio $M/N$.}
\label{fig:csmu_time}
\end{figure}

\textb{\Figref{csmu_time} shows the runtime of the algorithms in \Figref{csmu_iid}. 
Our implementation used MATLAB (R2015b) on an RHEL workstation with an 8-core Intel i7 processor. 
Although, for WSS-TLS, we used a grid-search to optimize $\lambda$ in \eqref{WSSTLS}, \Figref{csmu_time} only shows the runtime of WSS-TLS after $\lambda$ was chosen.
\Figref{csmu_time} shows BAd-VAMP running much faster than lifted VAMP and WSS-TLS, and slightly slower than PBiGAMP.}

\subsection{Self-Calibration}

In self-calibration \cite{Ling:IP:15}, the goal is to recover the $K$-sparse signal vector $\cb$ and the calibration parameters $\bbf$ from measurements of the form $\ybf = \Diag(\Hbf\bbf)\Psibf\cbf$ with known $\Hbf\in\Real^{M\times Q}$ and $\Psibf\in\Real^{M\times N}$.
Here, $\Hbf\bbf$ represents an unknown vector of gains on the measurements, where the gain vector is believed to lie in the $Q$-dimensional subspace spanned by the columns of $\Hbf$.
For our experiment, $M=128$, $N=256$, $\Psibf$ and $\bbf$ where drawn i.i.d.\ $\normal(0,1)$, 
$\Hbf$ was constructed using $Q$ randomly selected columns of the Hadamard matrix, and $\cb$ was drawn with uniformly random support and with $K$ non-zero elements from $\normal(\zero,\Ibf)$. 

\begin{figure}[t]
\centering
\newcommand{\sz}{0.7}
\psfrag{K}[t][t][\sz]{\sf sparsity $K$} 
\psfrag{Nb}[b][B][\sz]{\sf \# parameters $Q$} 
\psfrag{EM-P-BiG-AMP}[B][B][\sz]{\sf {EM-PBiGAMP}}
\psfrag{VAMP-Lift}[B][B][\sz]{\sf {VAMP-Lift}}
\psfrag{SparseLift}[B][B][\sz]{\sf {SparseLift}}
\psfrag{EM-VAMP}[B][B][\sz]{\sf {BAd-VAMP}}
\includegraphics[width=\figsize,trim=17mm 10mm 22mm 8mm,clip]{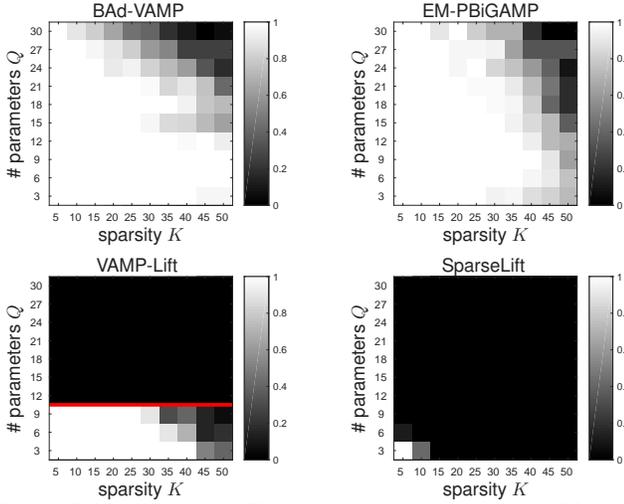}
\caption{Self-calibration: Empirical success rate (over 50 trials) for several algorithms versus number of calibration parameters $Q$ and sparsity $K$.  For computational reasons, VAMP-Lift was simulated only for $Q<10$.}
\label{fig:calibration}
\end{figure}

\Figref{calibration} shows the rate of successful recovery versus subspace dimension $Q$ and sparsity $K$ for several algorithms. 
A recovery $(\hat{\bbf},\hat{\cbf})$ was considered ``successful'' when $\|\widehat{\bbf}\widehat{\cbf}\tran-\bbf\cbf\tran\|_F^2/\|\bbf\cbf\tran\|_F^2$ $\leq-50$~dB.
From the figure, we see that the performance of BAd-VAMP is similar to that of EM-PBiGAMP, and even slightly better when $Q$ is small and $K$ is large.
Meanwhile, BAd-VAMP appears significantly better than both lifted VAMP and SparseLift from \cite{Ling:IP:15}.
SparseLift is a convex relaxation with provable guarantees \cite{Ling:IP:15}.
For computational reasons (recall the discussion in \secref{prior}), it was difficult to simulate lifted VAMP for $Q\geq 10$.

\blue
\subsection{Calibration in Tomography}

We consider the problem of reconstructing an image from a sequence of tomographic projections, where the projections along each direction are scaled by an unknown calibration gain.  
In particular, let $\Psibf_{\omega}$ be the tomographic projection matrix\footnote{\blue We used the matrix form of the Radon transform instead of the operator form to avoid numerical error when implementing the adjoint.}
(i.e., Radon transform) corresponding to angle $\omega\in[0,\pi]$. 
Our goal is to reconstruct the image $\xbf$ from measurements 
\begin{align}
\ybf
=\mat{b_1\Psibf_{\omega_1}\\[-1mm]\vdots\\b_K\Psibf_{\omega_K}}\xbf + 
\wbf \text{~~for~~} \wbf\sim \normal(\zero,\gamma_w^{-1}\Ibf)
\label{eq:yct} ,
\end{align}
where $b_k\sim\mc{N}(1,\sigma_b^2)$ are unknown.
Note that, by defining $\Abf_k=[\zero,\dotso,\Psibf_{\omega_k}\tran,\dotso,\zero]\tran$, we can write $\ybf=\Abf(\bbf)\xbf + \normal(\zero,\gamma_w^{-1}\Ibf)$ for $\Abf(\bbf) = \sum_{k=1}^K b_k\Abf_k$, which matches \eqref{bilinear3}. 

In an attempt to solve the above problem, we used BAd-VAMP to recover the image $\xbf$ while simultaneously learning the calibration gains $\bbf$.
For this, we used BAd-VAMP in ``plug-and-play'' mode, where the BM3D image denoiser \cite{Dabov:TIP:07} was used to implement the $\gbf_1(\cdot)$ function in \algref{EP}. 
Due to the inherent scaling ambiguity of the problem (i.e., if $(\hat{\xbf},\hat{\bbf})$ is a solution then so is $(\alpha\hat{\xbf},\alpha^{-1}\hat{\bbf})$ for any $\alpha>0$), we scaled the image estimate $\hat{\xbf}$ by the $\alpha$ that minimized $\|\xbf-\alpha\hat{\xbf}\|$ before computing the PSNR.

As baselines, we also tested the
VAMP \cite{Rangan:VAMP}, 
total variation (TV) \cite{Rudin:PhyD:92,Beck:TIP:09}
\blue
and 
regularization-by-denoising (RED) \cite{Romano:JIS:17,Reehorst:TCI:19}
approaches (see descriptions below).
These approaches all assume a noisy \emph{linear} data model of the form $\ybf=\hat{\Abf}\xbf+\mc{N}(0,\gamma_w^{-1}\Ibf)$ with known $\hat{\Abf}$.
To apply them to \eqref{yct}, we considered two cases:
the \emph{genie-calibrated} (GC) case, where a genie supplies the true gains $\bbf$ and the algorithm uses $\hat{\Abf}=\Abf(\bbf)$,
and the \emph{un-calibrated} (UC) case, where $\bbf$ is unknown and the algorithms assume $\hat{\Abf}=\Abf(\one)$.  
In the latter case, the $\hat{\Abf}$-based model is mismatched to the data-generation model \eqref{yct}. 
For fair comparison, we scaled the GC and UC image estimates $\hat{\xbf}$ by the $\alpha$ that minimized $\|\xbf-\alpha\xhat\|_2$ before computing the PSNR.

\begin{figure}[t]
\centering
    \begin{subfigure}[t]{0.15\textwidth}
        \centering
        \newcommand{\sz}{0.7}
        \psfrag{original}[b][b][\sz]{\sf Original}
        \includegraphics[scale=0.22,trim=5mm 0mm 5mm 0mm,clip]{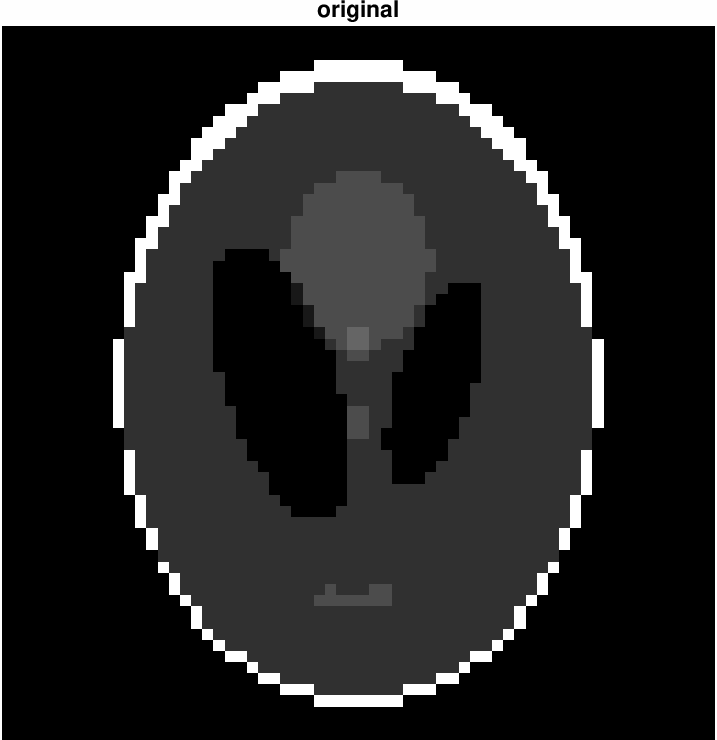}
    \end{subfigure}%
    ~
    \begin{subfigure}[t]{0.15\textwidth}
        \centering
        \newcommand{\sz}{0.7}
        \psfrag{BAdVAMP calibration: 38.54 (dB)}[b][b][\sz]{\sf BAdVAMP: 38.54}
        \includegraphics[scale=0.22,trim=5mm 0mm 5mm 0mm,clip]{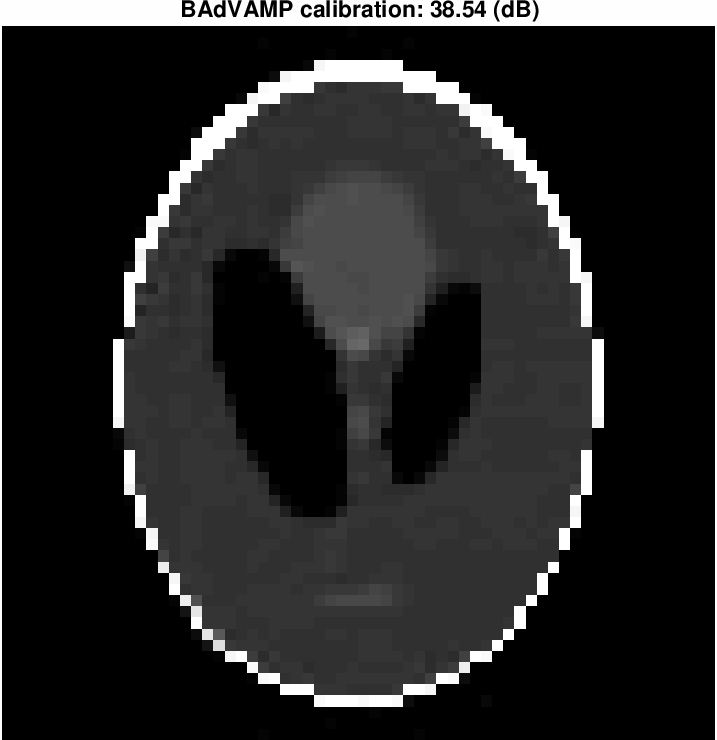}
    \end{subfigure}%
    ~
    \vspace{10pt}

    \begin{subfigure}[t]{0.15\textwidth}
        \centering
        \newcommand{\sz}{0.7}
        \psfrag{VAMP genie: 39.22 (dB)}[b][b][\sz]{\sf VAMP (GC): 39.22}
        \includegraphics[scale=0.22,trim=5mm 0mm 5mm 0mm,clip]{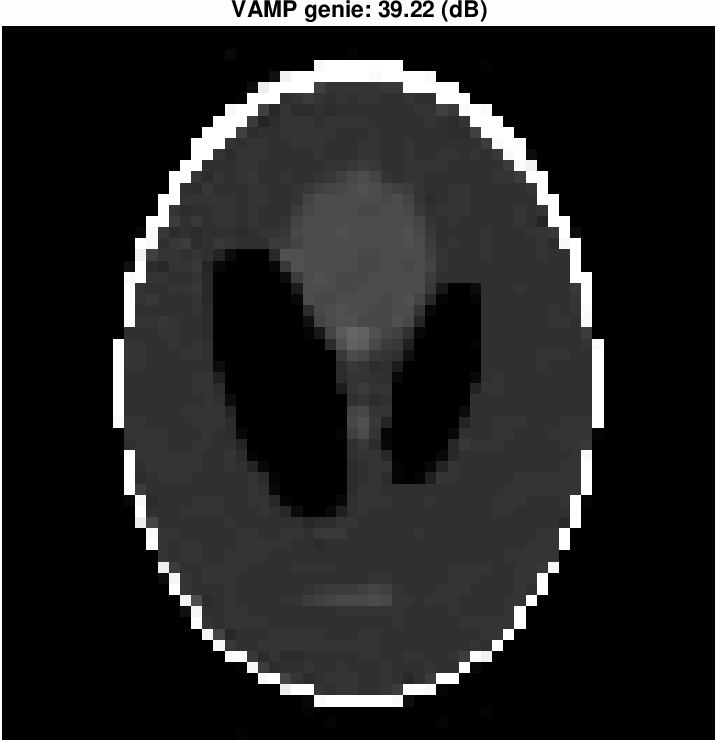}
    \end{subfigure}
    ~
    \begin{subfigure}[t]{0.15\textwidth}
        \centering
        \newcommand{\sz}{0.7}
        \psfrag{RED-ADMM genie: 35.79 (dB)}[b][b][\sz]{\sf RED (GC): 35.79}
        \includegraphics[scale=0.22,trim=5mm 0mm 5mm 0mm,clip]{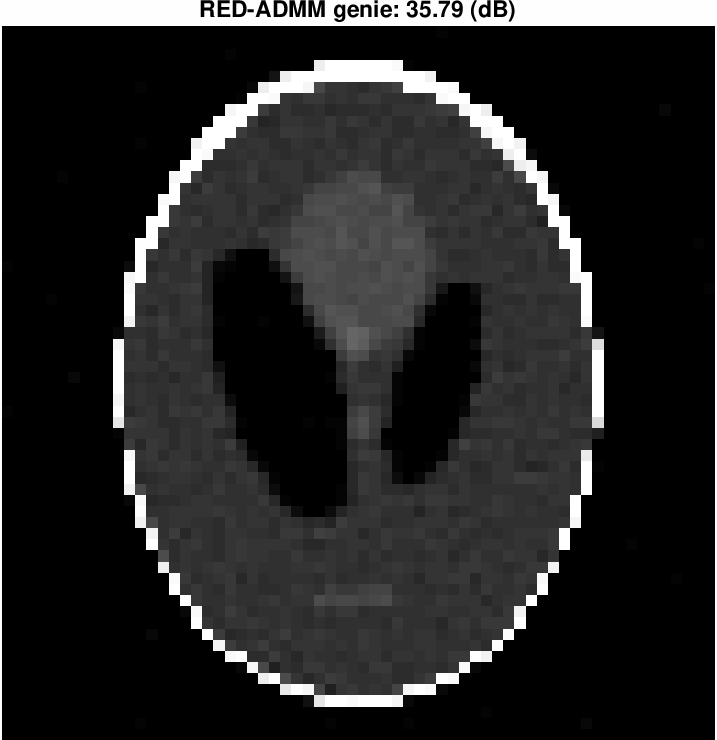}
    \end{subfigure}%
    ~
    \begin{subfigure}[t]{0.15\textwidth}
        \centering
        \newcommand{\sz}{0.7}
        \psfrag{FASTA genie: 31.87 (dB)}[b][b][\sz]{\sf TV (GC): 31.87}
        \includegraphics[scale=0.22,trim=5mm 0mm 5mm 0mm,clip]{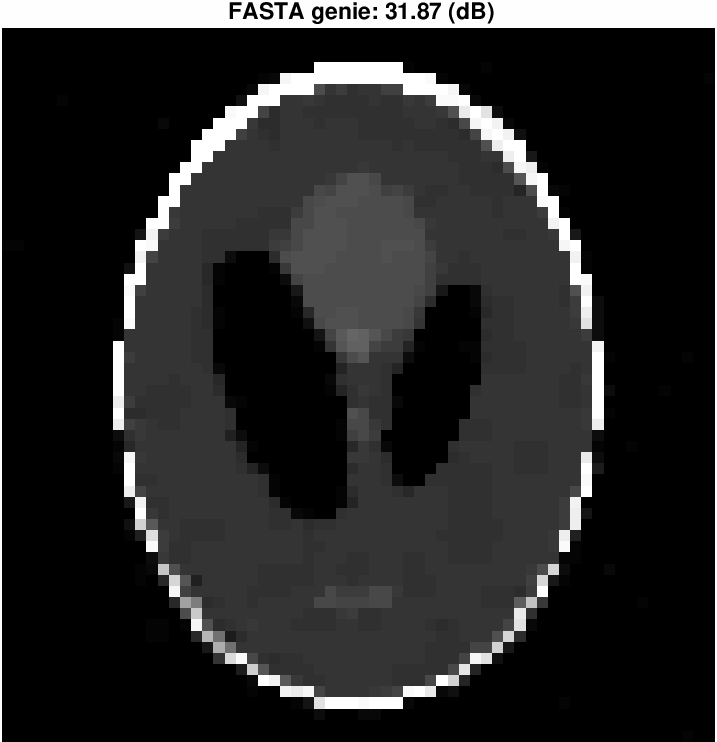}
    \end{subfigure}%
    \vspace{10pt}

    \begin{subfigure}[t]{0.15\textwidth}
        \centering
        \newcommand{\sz}{0.7}
        \psfrag{VAMP no calibration: 32.02 (dB)}[b][b][\sz]{\sf VAMP (UC): 32.02}
        \includegraphics[scale=0.22,trim=5mm 0mm 5mm 0mm,clip]{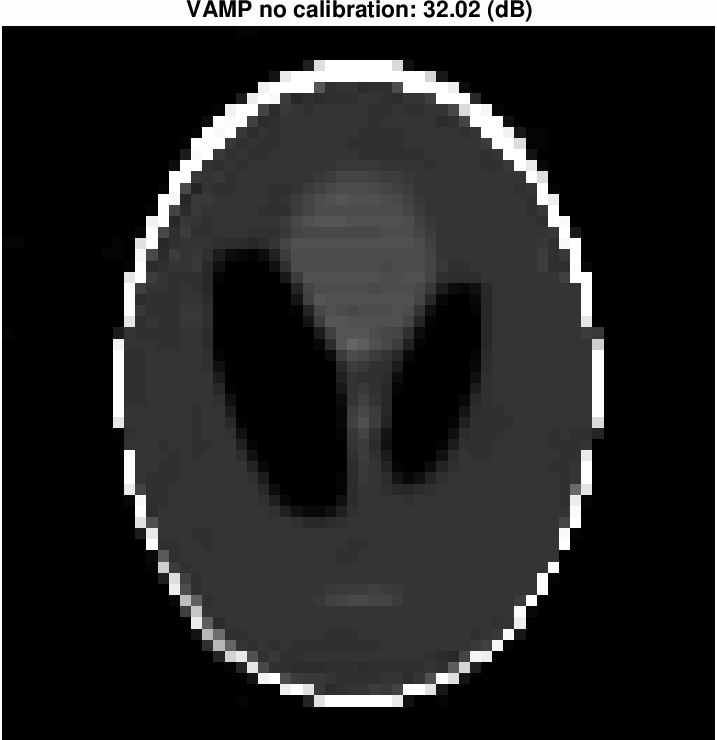}
    \end{subfigure}
    ~
    \begin{subfigure}[t]{0.15\textwidth}
        \centering
        \newcommand{\sz}{0.7}
        \psfrag{RED-ADMM no calibration: 30.11 (dB)}[b][b][\sz]{\sf RED (UC): 30.11}
        \includegraphics[scale=0.22,trim=5mm 0mm 5mm 0mm,clip]{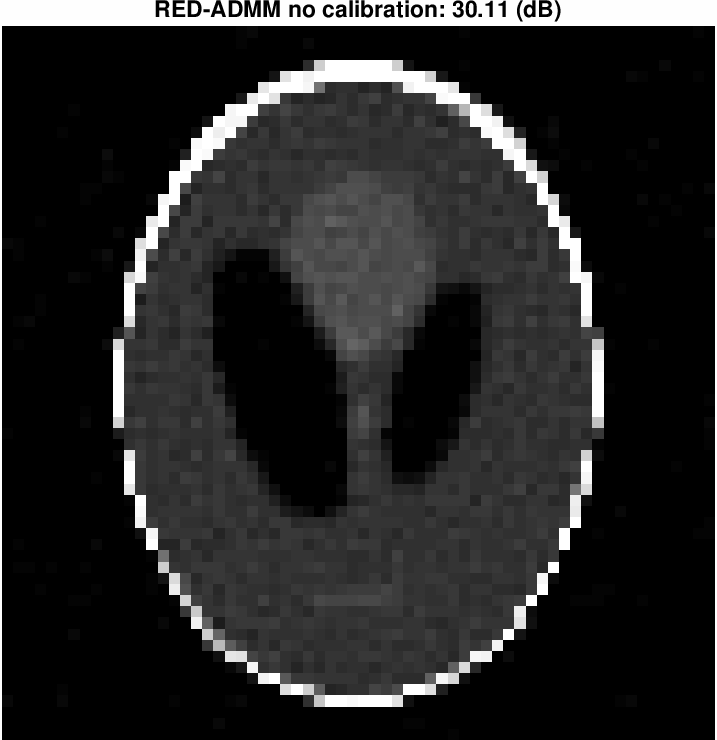}
    \end{subfigure}%
    ~
    \begin{subfigure}[t]{0.15\textwidth}
        \centering
        \newcommand{\sz}{0.7}
        \psfrag{FASTA no calibration: 25.22 (dB)}[b][b][\sz]{\sf TV (UC): 25.22}
        \includegraphics[scale=0.22,trim=5mm 0mm 5mm 0mm,clip]{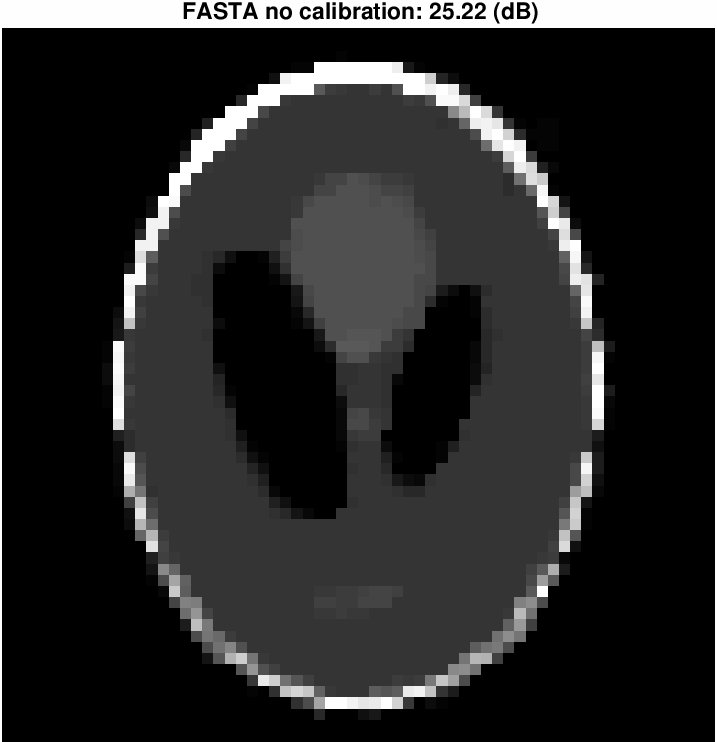}
    \end{subfigure}%
\caption{\textb{Calibration in tomography: Reconstruction PSNR (dB) of $64\times64$ Shepp-Logan phantom from $25$ equally spaced tomographic projections. In the genie-calibrated (GC) case, $\hat{\Abf}=\Abf(\bbf)$, while in the un-calibrated (UC) case, $\hat{\Abf}=\Abf(\one)$.}}
\label{fig:tomography_figures}
\end{figure}

\blue
We now provide additional details on the experimental setup.
For $\xbf$, we used the modified Shepp-Logan phantom of size $64\times 64$, shown in the top-left panel of \figref{tomography_figures}.
For $\Abf(\cdot)$, we used $K=25$ projections spaced uniformly in $\omega\in[0,\pi]$.
The calibration gains $\bbf$ were generated using $\sigma_b=0.06$, and
the noise precision $\gamma_w$ was set to achieve an SNR of 
$\Exp[\|\Abf(\bbf)\xbf\|^2]/\Exp[\|\wbf\|^2]=40$~dB.
The TV method \cite{Rudin:PhyD:92} computes
\begin{align}
\xhat = \arg\min_{\xbf} \bigg\{ \frac{1}{2}\|\ybf - \hat{\Abf}\xbf\|_2^2 +  \lambda_t\|\nabla\xbf\|_{2,1} \bigg\}
\label{eq:TV}
\end{align}
for the isotropic TV operator 
\begin{align}
\|\nabla \xbf\|_{2,1} &= \sum_{i,j} \sqrt{(x_{i,j}-x_{i,j-1})^2 + (x_{i,j}-x_{i-1,j})^2} .
\end{align}
We solved \eqref{TV} using FASTA \cite{Goldstein:14},
and tuned $\lambda_t$ to maximize the PSNR.
\blue
RED \cite{Romano:JIS:17,Reehorst:TCI:19} solves the fixed-point equation 
\begin{align}
\hat{\Abf}\herm(\hat{\Abf}\hat{\xbf}-\ybf) + \lambda_r(\hat{\xbf} - \rhobf(\hat{\xbf},\tau))
= \zero
\label{eq:red}
\end{align}
for $\hat{\xbf}$, where $\rhobf(\cdot,\tau)$ is an image denoising algorithm with noise-variance $\tau$.
For our experiment, 
we used BM3D for $\rhobf(\cdot,\tau)$,
solved \eqref{red} using the ADMM method from \cite{Romano:JIS:17} with 200 iterations, 
and tuned both $\lambda_r$ and $\tau$ to maximize PSNR.
For BAd-VAMP, we initialized $\bbf$ to $\one$, used damping $\zeta=0.1$, assumed known noise precision $\gamma_w$, and used at most $100$ iterations. 
For VAMP, we 
initialized $\gamma_1^0=10^{-4}$ and used at most $100$ iterations.

\begin{figure}
\centering
    \newcommand{\sca}{0.20}
    \begin{subfigure}[t]{0.15\textwidth}
        \centering
        \newcommand{\sz}{0.7}
        \psfrag{BAdVAMP error}[b][b][\sz]{\sf BAdVAMP}
        \includegraphics[scale=\sca,trim=5mm 0mm 0mm 0mm,clip]{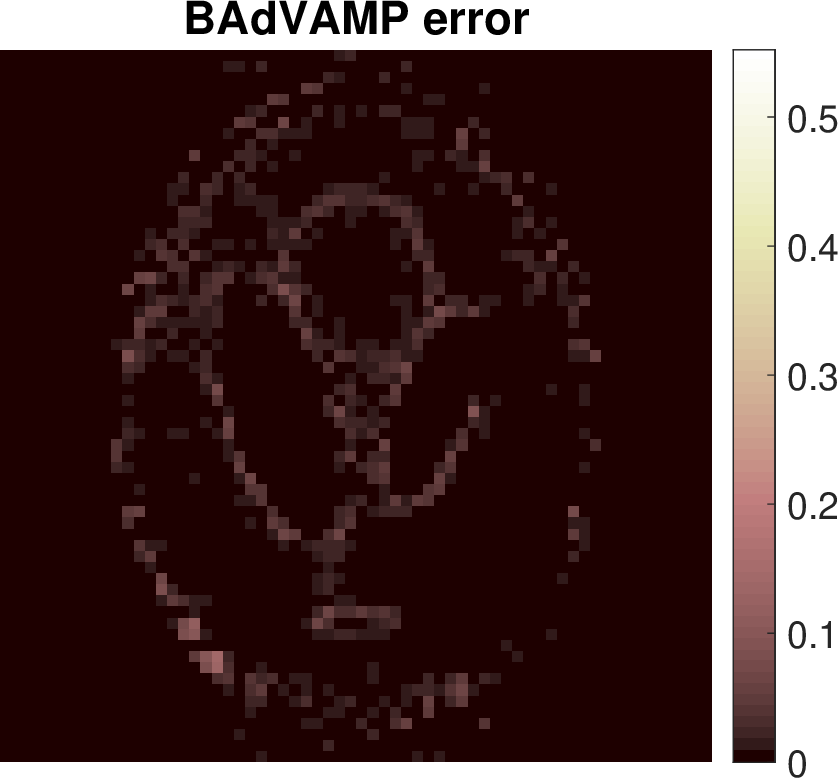}
    \end{subfigure}%
    ~
    \vspace{10pt}

    \begin{subfigure}[t]{0.15\textwidth}
        \centering
        \newcommand{\sz}{0.7}
        \psfrag{VAMP genie error}[b][b][\sz]{\sf VAMP (GC)}
        \includegraphics[scale=\sca,trim=5mm 0mm 0mm 0mm,clip]{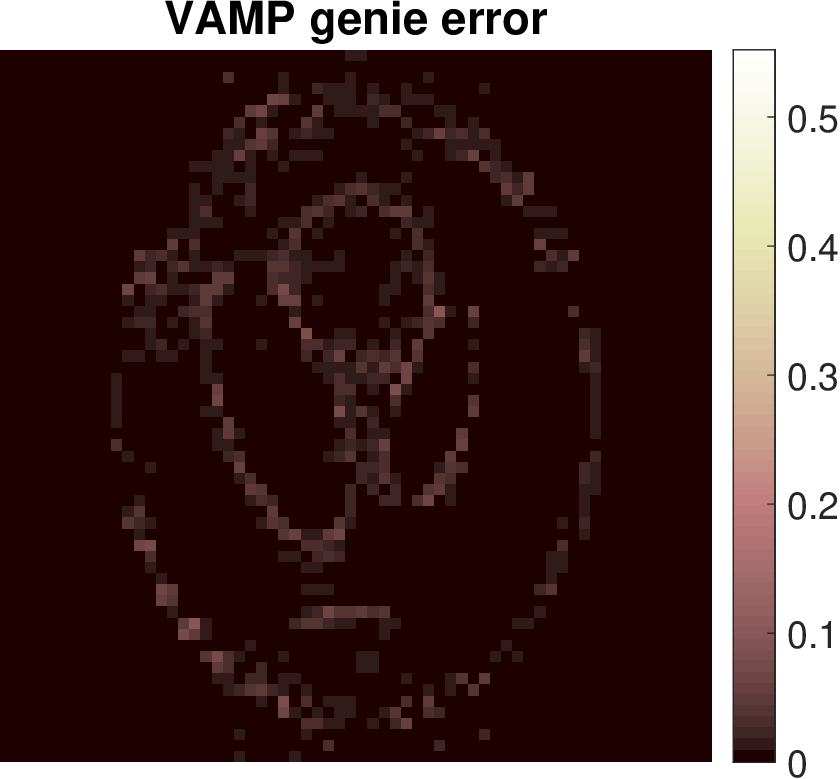}
    \end{subfigure}
    ~
    \begin{subfigure}[t]{0.15\textwidth}
        \centering
        \newcommand{\sz}{0.7}
        \psfrag{RED-ADMM genie error}[b][b][\sz]{\sf RED (GC)}
        \includegraphics[scale=\sca,trim=5mm 0mm 0mm 0mm,clip]{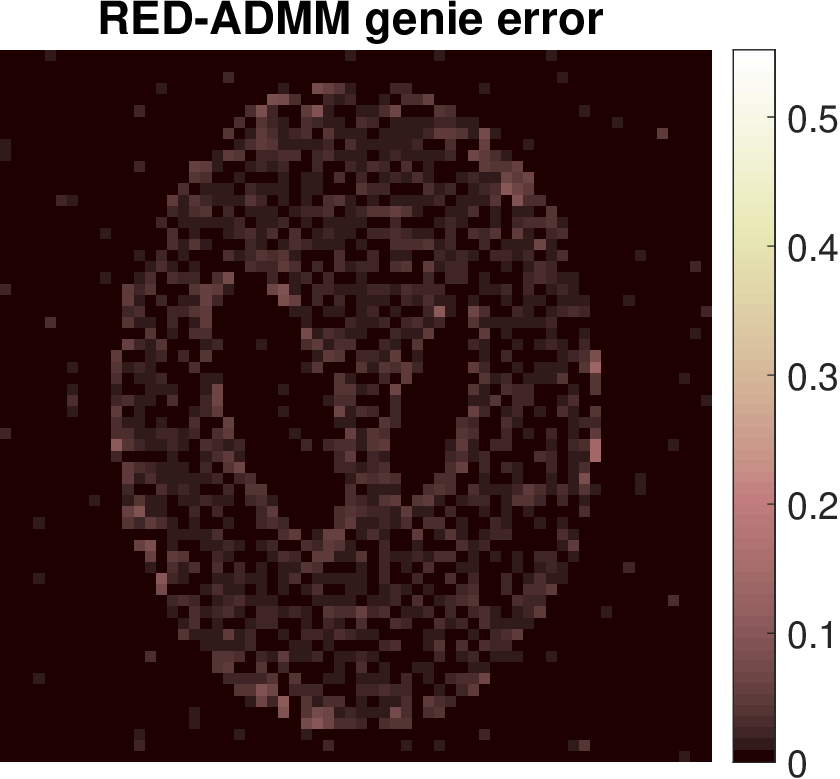}
    \end{subfigure}%
    ~~
    \begin{subfigure}[t]{0.15\textwidth}
        \centering
        \newcommand{\sz}{0.7}
        \psfrag{FASTA genie error}[b][b][\sz]{\sf TV (GC)}
        \includegraphics[scale=\sca,trim=5mm 0mm 0mm 0mm,clip]{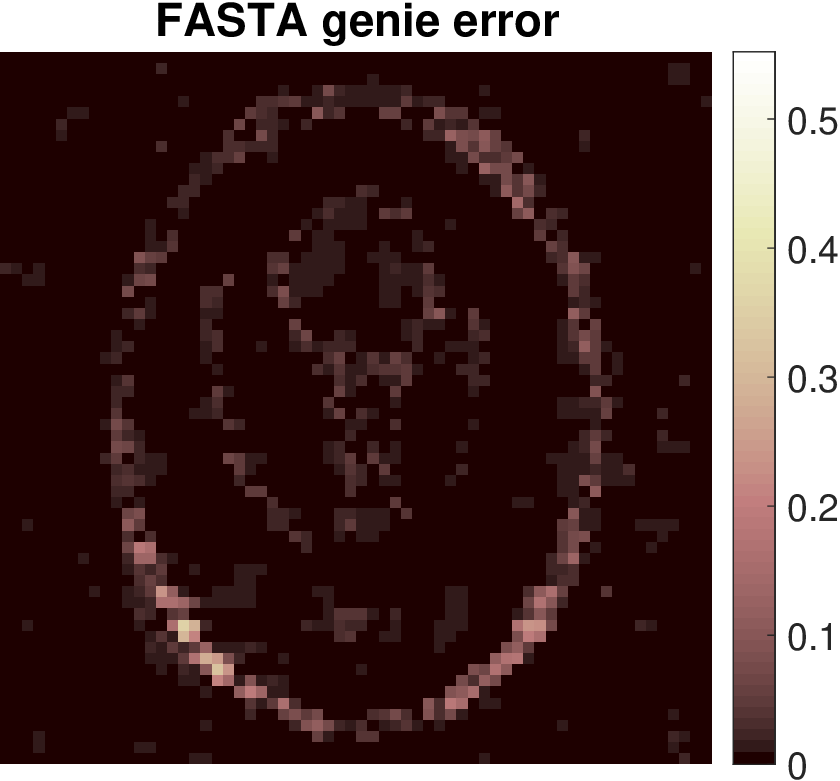}
    \end{subfigure}%
    \vspace{10pt}

    \begin{subfigure}[t]{0.15\textwidth}
        \centering
        \newcommand{\sz}{0.7}
        \psfrag{VAMP nocalib error}[b][b][\sz]{\sf VAMP (UC)}
        \includegraphics[scale=\sca,trim=5mm 0mm 0mm 0mm,clip]{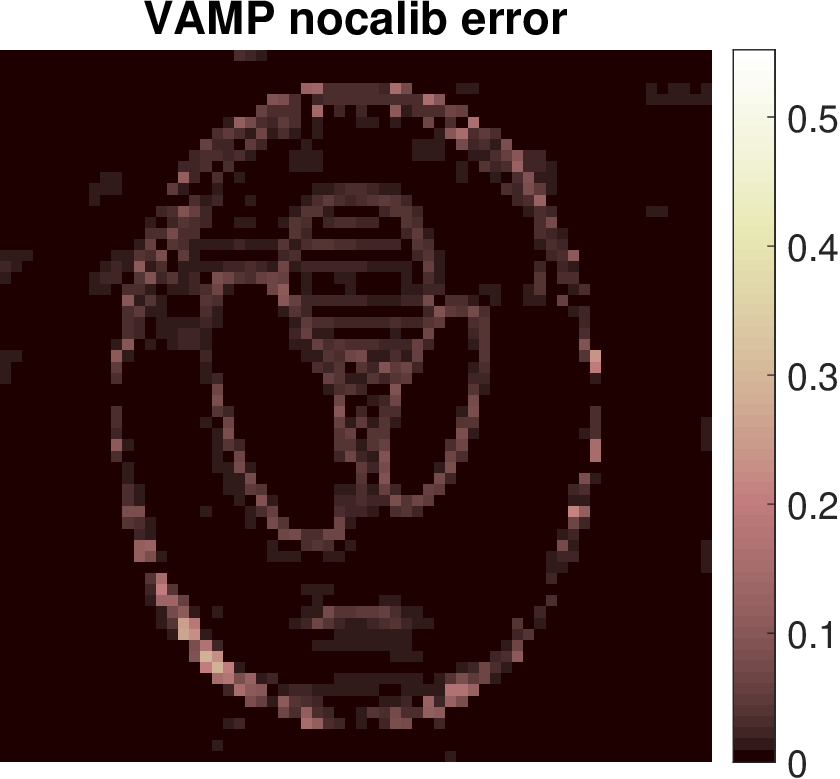}
    \end{subfigure}
    ~
    \begin{subfigure}[t]{0.15\textwidth}
        \centering
        \newcommand{\sz}{0.7}
        \psfrag{RED-ADMM nocalib error}[b][b][\sz]{\sf RED (UC)}
        \includegraphics[scale=\sca,trim=5mm 0mm 0mm 0mm,clip]{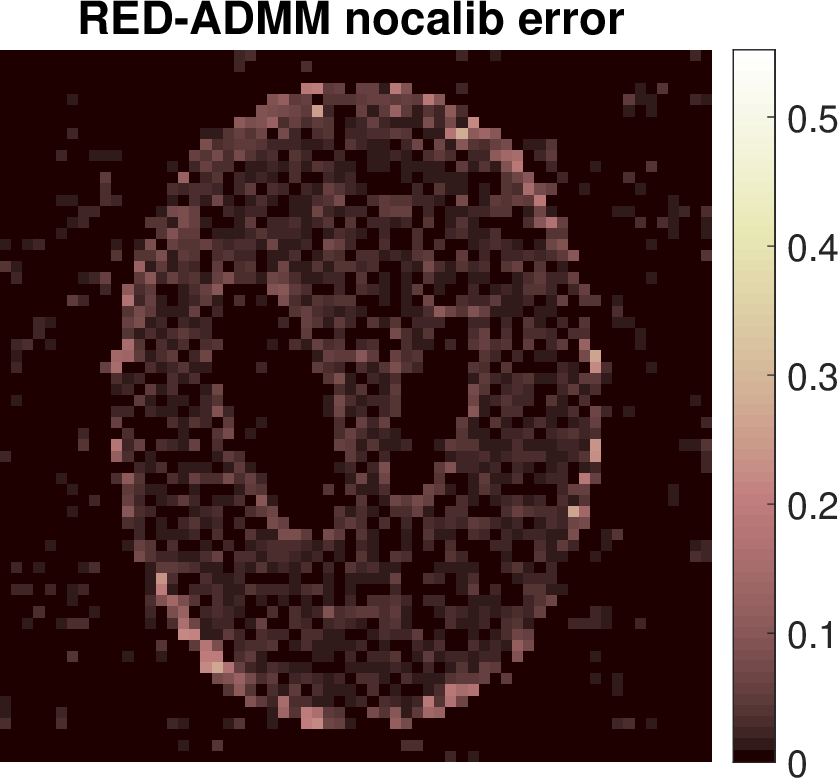}
    \end{subfigure}%
    ~~
    \begin{subfigure}[t]{0.15\textwidth}
        \centering
        \newcommand{\sz}{0.7}
        \psfrag{FASTA nocalib error}[b][b][\sz]{\sf TV (UC)}
        \includegraphics[scale=\sca,trim=5mm 0mm 0mm 0mm,clip]{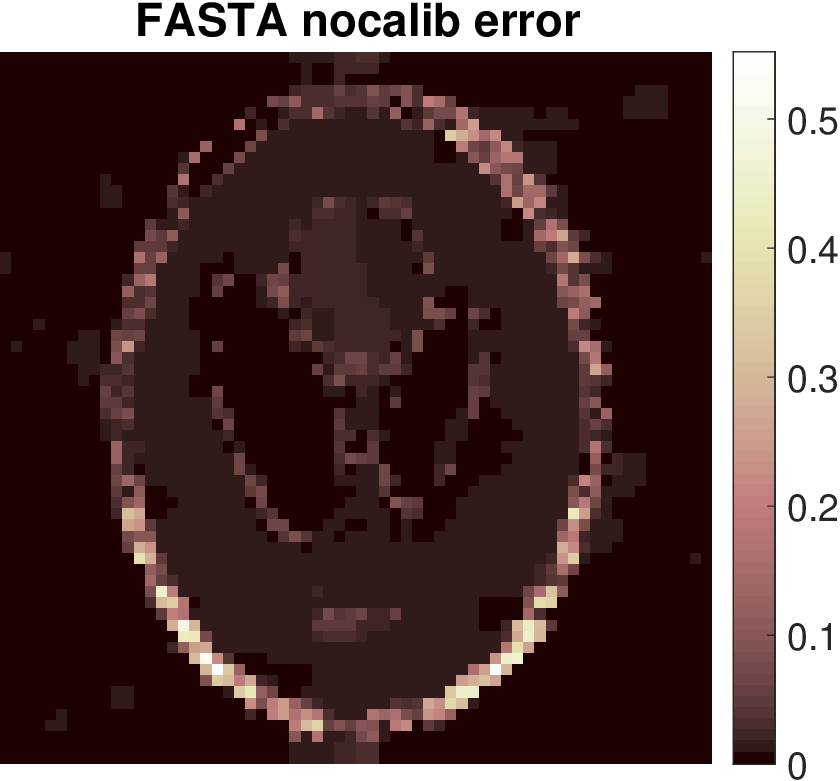}
    \end{subfigure}%
\caption{\textb{Calibration in tomography: Error images for the reconstructions shown in \figref{tomography_figures}. }}
\label{fig:tomography_error_figures}
\end{figure}

\tabref{tomography_table} reports the median PSNR achieved by each algorithm across $10$ random draws of $\bbf$ and $\wbf$, 
\figref{tomography_figures} shows example image recoveries,
and \figref{tomography_error_figures} shows the corresponding error images.
From \tabref{tomography_table}, we see that the PSNR performance of BAd-VAMP (which does not know $\bbf$) is nearly as good as genie-calibrated VAMP, and $1.7$~dB better than genie-calibrated RED.
Furthermore, the PSNR performance of BAd-VAMP is more than $6.6$~dB better than un-calibrated VAMP and $7$~dB better than un-calibrated RED. 
The uncalibrated VAMP, RED, and TV recoveries in \figref{tomography_figures} are plagued by either streaking artifacts and/or loss of detail (e.g., note the disappearance of the small white dots in uncalibrated TV).
But the BAdVAMP image recovery in \figref{tomography_figures} shows no streaking artifacts and a high level of detail.
Likewise, \figref{tomography_error_figures} shows that TV has trouble correctly recovering the white outer ellipse, RED has trouble in the interior region, but BAd-VAMP does well throughout.


\begin{table}[t]\blue
\begin{center}
    \caption{\blue PSNR (dB) in the tomography experiment}
    \label{tab:tomography_table}
    \begin{tabular}{|c||c|c|c|c|} 
        \hline
        measurements & BAd-VAMP & VAMP & RED & TV \\ 
        \hline
        genie calibrated (GC) & --- 
                & \textbf{39.57} & 36.56 & 33.27 \\ 
        \hline
        un-calibrated (UC) & \textbf{38.27} & 31.62 & 31.24 & 26.79 \\
         \hline
    \end{tabular}
\end{center}
\end{table}
  
\black
\subsection{Noiseless Dictionary Learning}

In dictionary learning (DL) \cite{Rubinstein:PROC:10}, the goal is to find a dictionary matrix $\Ab\in\Real^{M\times N}$ and a sparse matrix $\Xb\in\Real^{N\times L}$ such that a given matrix $\Yb\in\Real^{M\times L}$ can be approximately factored as $\Yb\approx\Ab\Xb$. 
In this section, we test the proposed BAd-VAMP algorithm for DL by generating $\Yb=\Ab\Xb$ such that $\Xb$ has $K$-sparse columns, and measuring the NMSE on the resulting estimates of $\Ab$ and $\Xb$.

We consider two cases: i) where the true $\Ab$ is structured as $\Ab=\sum_{i=1}^Q b_i\Ab_i$ with known $\{\Ab_i\}_{i=1}^Q$ (recall \eqref{A_affine}),
and ii) where the true $\Ab$ is unstructured (recall \eqref{A_unstructured}).
In either case, the pair $(\Ab,\Xb)$ is recoverable only up to an ambiguity: a scalar ambiguity in the structured case and a generalized permutation ambiguity in the unstructured case.
Thus, when measuring reconstruction quality, we consider
\begin{align}
\text{NMSE}(\hat{\Ab}) &\defn \min_{\lambda\in\Real}\frac{\|\Ab-\lambda\hat{\Ab}\|^2_F}{\|\Ab\|^2_F} 
\end{align}
in the structured case and
\begin{align}
\text{NMSE}(\hat{\Ab}) &\defn \min_{\Pb\in\mathcal{P}}\frac{\|\Ab-\hat{\Ab}\Pb\|^2_F}{\|\Ab\|^2_F}
\label{eq:NMSE_A_unstructured}
\end{align}
in the unstructured case, where $\mathcal{P}$ denotes the set of generalized permutation matrices.\footnote{If $\Pbf$ is a generalized permutation matrix then $\Pbf=\bm{\Pi}\Dbf$, where $\bm{\Pi}$ is a permutation matrix and $\Dbf$ is a diagonal matrix.} 
For our experiments, we drew the coefficients of $\{\Ab_i\}_{i=1}^Q$ and $\bbf$ as i.i.d.\ $\normal(0,1)$ with $Q=N$ in the structured case, and we drew the coefficients of $\Ab$ as i.i.d.\ $\normal(0,1)$ in the unstructured case.

\begin{figure}[t]
\centering
    \begin{subfigure}[t]{0.5\textwidth}
        \centering
        \newcommand{\sz}{0.7}
        \psfrag{a}[b][b][\sz]{\sf unstructured}
        \psfrag{b}[b][b][\sz]{\sf structured}
        \psfrag{N}[t][t][\sz]{\sf dimension $N$}
        \psfrag{L}[b][B][\sz]{\sf training length $L$}
        \includegraphics[width=3.1in,height=1.28in,trim=25mm 2mm 28mm 2mm,clip]{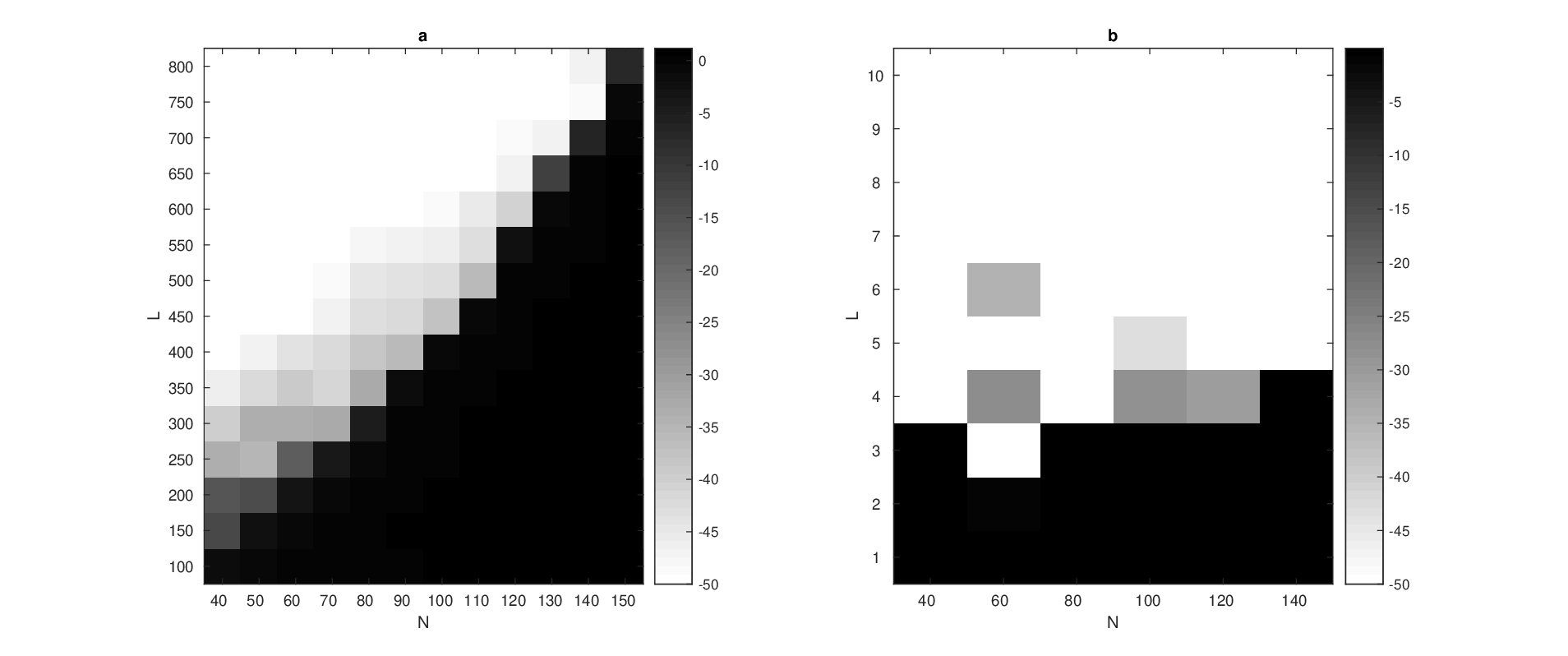}
     \end{subfigure}%
     \vspace{5pt}

    \begin{subfigure}[t]{0.5\textwidth}
        \centering
        \newcommand{\sz}{0.7}
        \psfrag{N}[t][t][\sz]{\sf dimension $N$}
        \psfrag{y1}[b][B][\sz]{\sf NMSE($\hat{\Abf}$) [dB]}
        \includegraphics[width=2.9in,trim=5mm 5mm 5mm 5mm,clip]{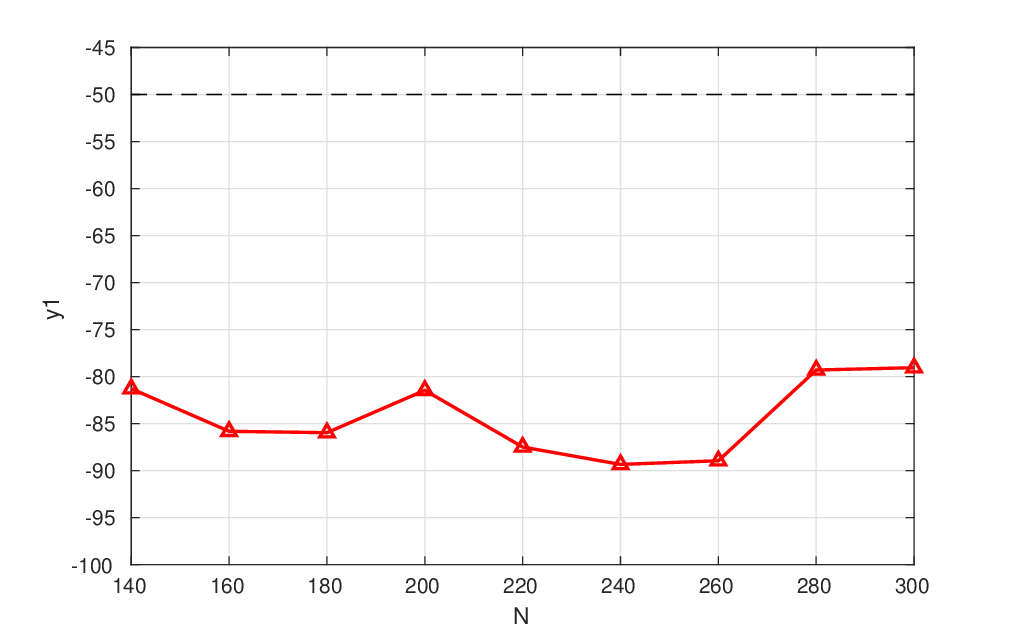}
    \end{subfigure}
    \caption{\textb{BAd-VAMP dictionary learning with $N\times N$ dictionary $\Abf$ and $N\times L$ code matrix $\Xbf$ with sparsity rate $K/N=0.2$.
    Top left: success-rate for unstructured dictionary versus $N$ and $L$. 
    Top right: success-rate for structured dictionary with $Q=N$ free parameters. 
    Bottom: Median $\text{NMSE}(\hat{\Ab})$ over 20 trials versus $N$ for an unstructured dictionary with $L=6N\ln N$.}}
\label{fig:NL_iid_linear}
\end{figure}

In our first experiment, we fixed the sparsity rate at $K/N=0.2$ and we varied both the dictionary dimension $N$ and the training length $L$.
\textb{The top-right panel of} \figref{NL_iid_linear} suggests that, as the dimension $N$ grows, a \emph{fixed} training length $L$ is sufficient to successfully recover $\Abf$ in the structured case with $Q=N$. 
By ``successfully recover,'' we mean that $\text{NMSE}(\hat{\Abf})\leq-50$~dB.
Note that this latter prescription for $L$ is consistent with the theoretical analysis in \cite{Spielman:COLT:12}.
\textb{In the unstructured case, the bottom panel of \Figref{NL_iid_linear} shows the median $\text{NMSE}(\hat{\Abf})$ versys $N$ when $L=6N\ln N$. 
Together, the top-left and bottom panels of \figref{NL_iid_linear} suggest that a training length of $L=O(N\ln N)$ suffices to successfully recover $\Abf$.}

%

\begin{figure}[t]
\centering
\newcommand{\sz}{0.7}
\psfrag{Dictionary Size}[t][t][\sz]{\sf dimension $N$} 
\psfrag{Sparsity}[b][B][\sz]{\sf sparsity $K$} 
\psfrag{K-SVD}[B][b][\sz]{\sf \raisebox{-0.8mm}{K-SVD}} 
\includegraphics[width=1.38in,height=1.31in,clip]{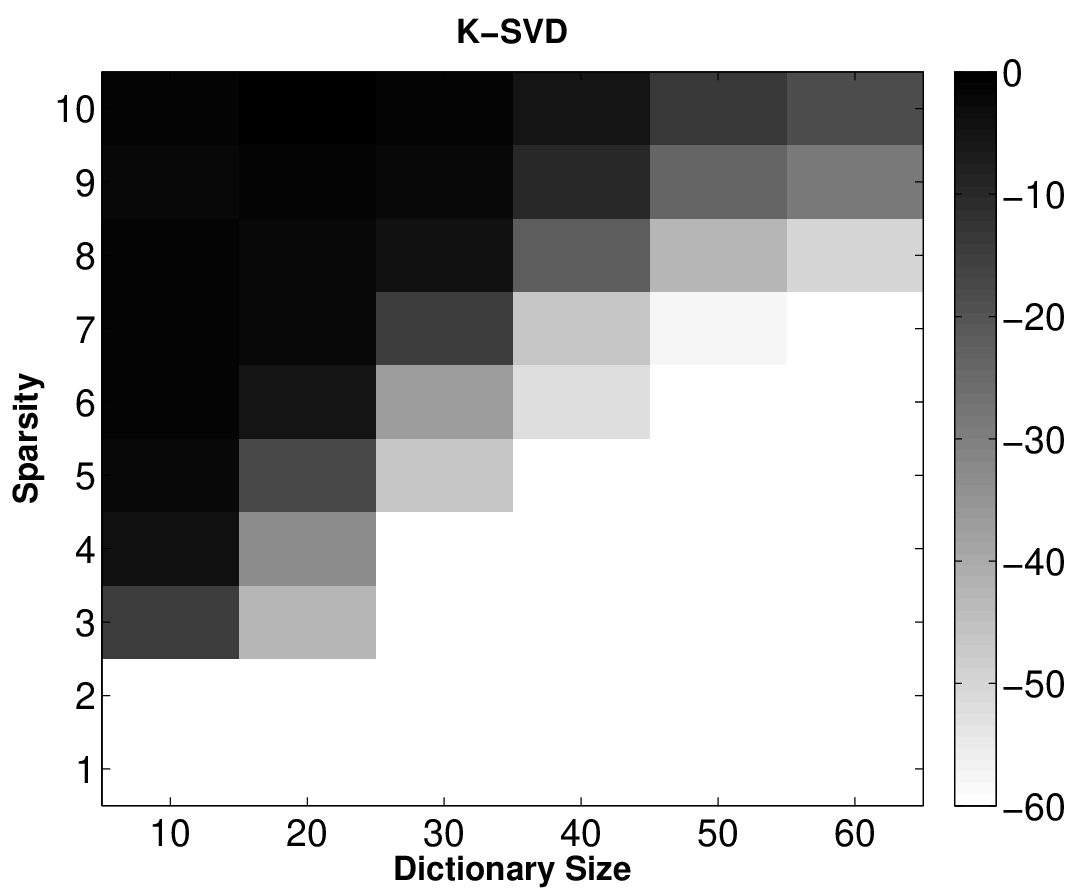}
\hspace{4.8mm}
\psfrag{SPAMS}[B][b][\sz]{\sf \raisebox{-0.8mm}{SPAMS}} 
\includegraphics[width=1.38in,height=1.31in,clip]{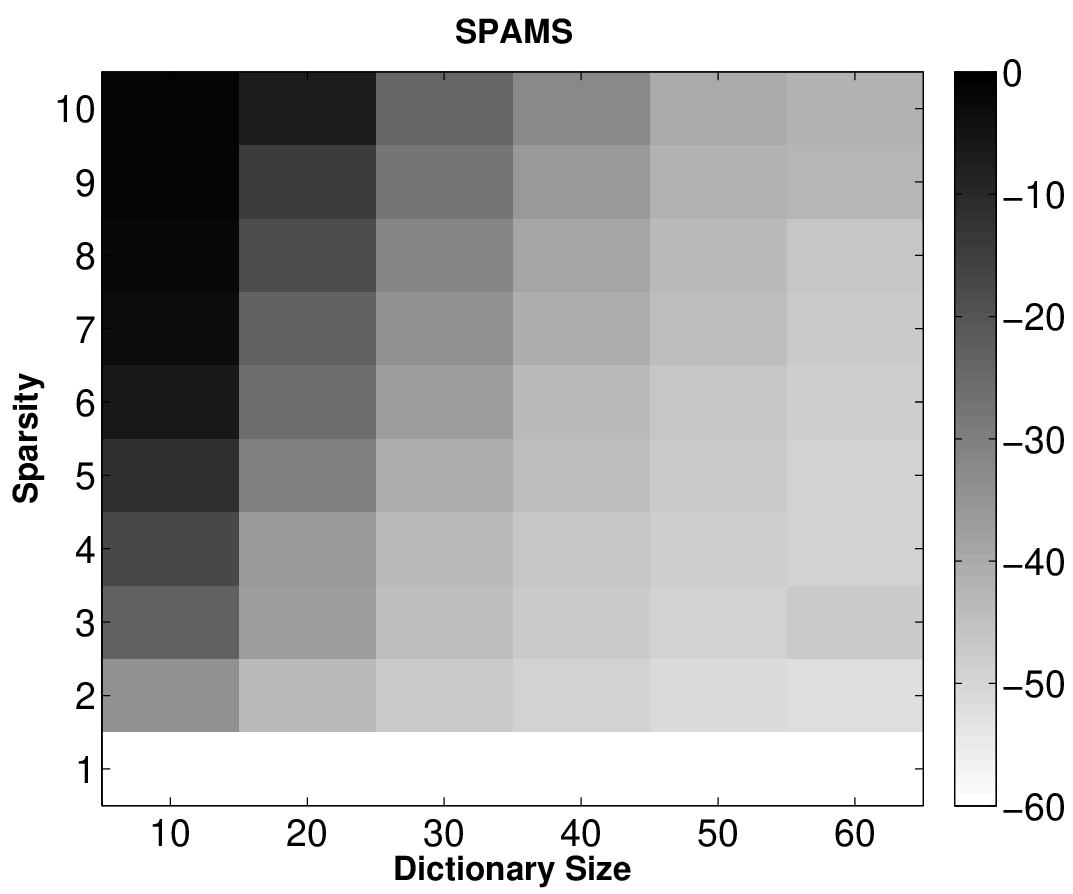}\\[4mm]
\psfrag{M}[t][t][\sz]{\sf dimension $N$} 
\psfrag{K}[b][B][\sz]{\sf sparsity $K$} 
\psfrag{NMSE(M,K)}[b][b][\sz]{\sf \raisebox{0.2mm}{BAd-VAMP}} 
\includegraphics[width=1.35in,height=1.25in,trim=6mm 2mm 12mm 5mm,clip]{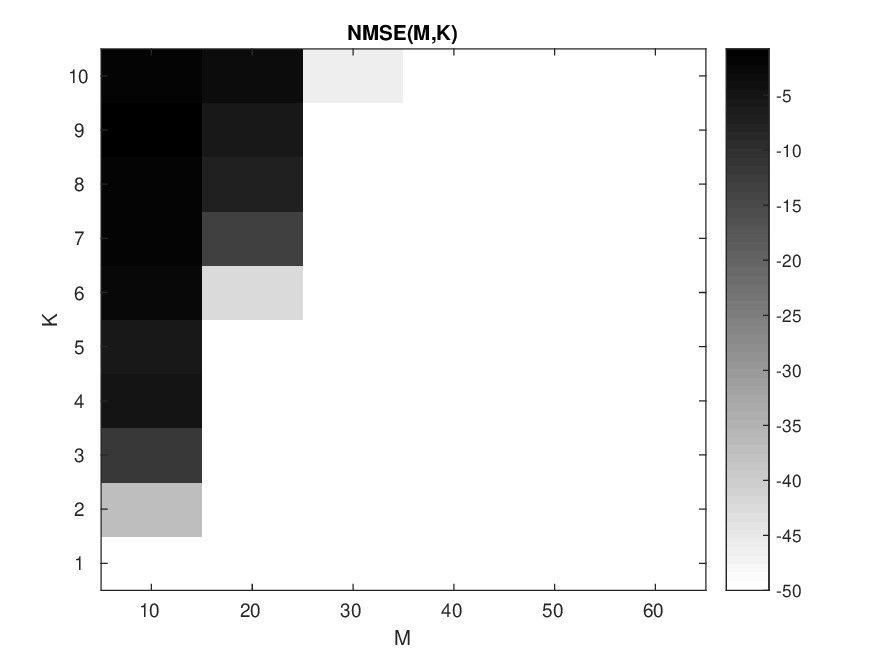}
\hspace{5mm}
\psfrag{EM-BiG-AMP}[B][b][\sz]{\sf \raisebox{-0.8mm}{EM-BiGAMP}} 
\includegraphics[width=1.38in,height=1.31in,clip]{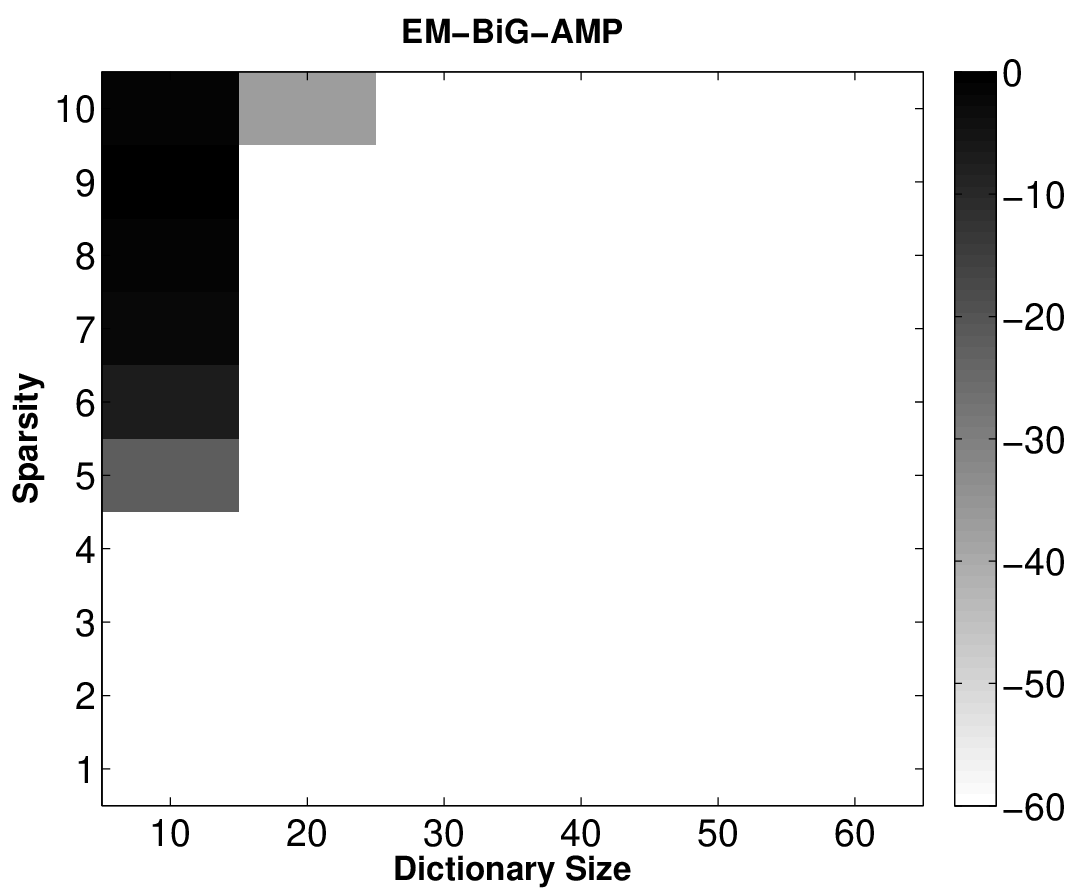}
\caption{Noiseless dictionary learning: Median NMSE$(\hat{\Ab})$ in dB (over 20 trials) versus dictionary dimension $N$ and sparsity $K$ with unstructured dictionary $\Ab\in\Real^{N\times N}$ and training length $L=5N\ln N$.}
\label{fig:MK_iid}
\end{figure}

In our second experiment, we focused on the unstructured case, fixed the training length at $L=5N\ln N$, and varied both the dictionary dimension $N$ and the sparsity $K$ in the columns of $\Xb$.
\Figref{MK_iid} shows that BAd-VAMP performed similarly to EM-BiGAMP~\cite{Parker:TSP:14b} for all but very small $N$, and much better than K-SVD \cite{Aharon:TSP:06} and SPAMS \cite{Mairal:JMLR:10}. 
The advantage of BAd-VAMP over EM-BiGAMP for DL will be illustrated in the sequel.

\subsection{Noisy, Ill-Conditioned Dictionary Learning}

In this section, we show the robustness of BAd-VAMP over EM-BiGAMP~\cite{Parker:TSP:14a} when learning ill-conditioned dictionaries from noisy measurements. 
To do so, we generated the measurements as $\Ybf=\Abf\Xbf+\Wbf$ and tested the algorithms in recovering $\Abf$ and $\Xbf$ (up to appropriate ambiguities).
The elements of $\Wbf$ were drawn i.i.d.\ $\normal(0,1/\gamma_w)$ with 
$\gamma_w$ chosen to achieve $\text{SNR}\defn\Exp[\|\Ab\Xbf\|_F^2]/\Exp[\|\Wbf\|_F^2]=40$~dB.
The true dictionary was generated as $\Ab=\Ub\Diag(\sb)\Vb\tran$, where $\Ub$ and $\Vb$ were drawn uniformly over the group of orthogonal matrices, and where the singular values in $\sb$ were chosen so that $s_i/s_{i-1}=\rho~\forall i$. 
The values of $s_0$ and $\rho$ were selected to obtain a desired condition number $\kappa(\Ab)$ while also ensuring $\|\Ab\|_F^2=N$.

\begin{figure}[t]
\centering
\newcommand{\sz}{0.7}
\psfrag{x1}[t][t][\sz]{\sf condition number $\kappa(\Ab)$} 
\psfrag{x2}[t][t][\sz]{\sf condition number $\kappa(\Ab)$} 
\psfrag{y1}[b][B][\sz]{\sf NMSE($\hat{\Ab}$) [dB]} 
\psfrag{y2}[b][B][\sz]{\sf NMSE($\hat{\Xb}$) [dB]}
\psfrag{EM-BiG-AMP}[l][l][0.4]{\sf \hspace{-0.25mm}EM-BiGAMP}
\psfrag{EM-VAMP}[l][l][0.4]{\sf \hspace{-0.25mm}BAd-VAMP}
\includegraphics[width=\figsize,trim=15mm 5mm 15mm 7mm,clip]{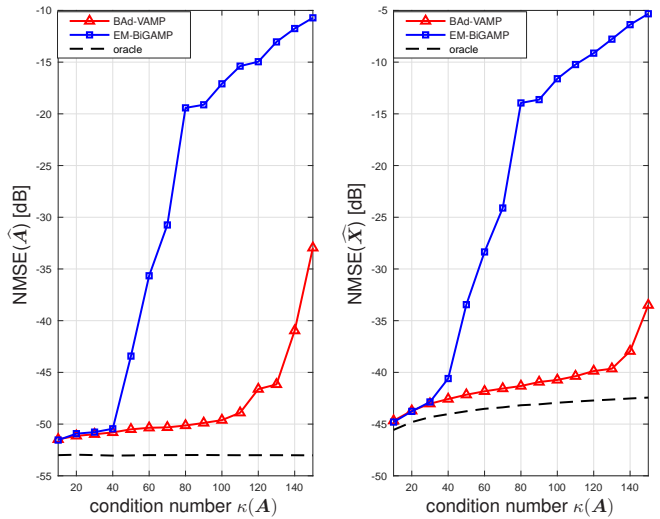}
\caption{Noisy dictionary learning: Median NMSE in dB (over 50 trials) versus condition number $\kappa(\Ab)$ for unstructured $\Ab\in\Real^{N\times N}$ with $N=64$, sparsity $K=13$, training length $L=5N\ln N$, and SNR $=40$ dB.}
\label{fig:dl_condA}
\end{figure}

\Figref{dl_condA} reports median $\text{NMSE}(\hat{\Ab})$ and $\text{NMSE}(\hat{\Xb})$ versus condition number $\kappa(\Abf)$ for the recovery of $\Ab\in\Real^{N\times N}$ and $K$-sparse $\Xbf\in\Real^{N\times L}$ from noisy measurements $\Ybf$.
For this figure, we used $K=13$, $N=64$, $L=5 N\ln N$, the unstructured definition of $\text{NMSE}(\hat{\Ab})$ from \eqref{NMSE_A_unstructured}, and a similar definition for $\text{NMSE}(\hat{\Xb})$.
In addition to showing the performance of BAd-VAMP and EM-PBiGAMP, the figure shows the performance of the known-$\Xbf$ oracle for the estimation of $\Abf$, as well as the known-$\Abf$ and known-support oracle for the estimation of $\Xbf$.
\Figref{dl_condA} shows that EM-BiGAMP gave near-oracle NMSE for $\kappa(\Ab)\leq 40$, but its performance degraded significantly for larger $\kappa(\Ab)$. 
In contrast, BAd-VAMP gave near-oracle NMSE for $\kappa(\Ab)\leq 110$, which suggests increased robustness to ill-conditioned dictionaries $\Ab$.

\section{Conclusion}

In this paper, we considered the problem of jointly recovering the vector $\bbf$ and the matrix $\Cbf$ from noisy measurements $\Ybf = \Abf(\bbf)\Cbf + \Wbf$, where $\Abf(\cdot)$ is a known affine linear function of $\bbf$ (i.e., $\Abf(\bbf)=\Abf_0+\sum_{i=1}^Q b_i \Abf_i$ with known matrices $\Abf_i$).
To solve this problem, we proposed the BAd-VAMP algorithm, which combines the VAMP algorithm \cite{Rangan:VAMP}, the EM algorithm \cite{Neal:Jordan:98}, and variance auto-tuning \cite{Kamilov:TIT:14} in a manner appropriate for bilinear recovery. 
We demonstrated numerically that the proposed approach has robustness advantages over other state-of-the-art bilinear recovery algorithms, including lifted VAMP \cite{Fletcher:NIPS:18} and EM-PBiGAMP \cite{Parker:JSTSP:16}.
As future work, we plan to rigorously analyze BAd-VAMP through the state-evolution formalism.



\bibliographystyle{IEEEtran}
\bibliography{macros_abbrev,phase,books,misc,comm,multicarrier,sparse,machine}

\begin{thebibliography}{10}
\providecommand{\url}[1]{#1}
\csname url@samestyle\endcsname
\providecommand{\newblock}{\relax}
\providecommand{\bibinfo}[2]{#2}
\providecommand{\BIBentrySTDinterwordspacing}{\spaceskip=0pt\relax}
\providecommand{\BIBentryALTinterwordstretchfactor}{4}
\providecommand{\BIBentryALTinterwordspacing}{\spaceskip=\fontdimen2\font plus
\BIBentryALTinterwordstretchfactor\fontdimen3\font minus
  \fontdimen4\font\relax}
\providecommand{\BIBforeignlanguage}[2]{{%
\expandafter\ifx\csname l@#1\endcsname\relax
\typeout{** WARNING: IEEEtran.bst: No hyphenation pattern has been}%
\typeout{** loaded for the language `#1'. Using the pattern for}%
\typeout{** the default language instead.}%
\else
\language=\csname l@#1\endcsname
\fi
#2}}
\providecommand{\BIBdecl}{\relax}
\BIBdecl

\bibitem{Candes:PROC:10}
E.~J. Cand{\`e}s and Y.~Plan, ``Matrix completion with noise,'' \emph{Proc.
  IEEE}, vol.~98, no.~6, pp. 925--936, Jun. 2010.

\bibitem{Candes:JACM:11}
E.~J. Cand{\`e}s, X.~Li, Y.~Ma, and J.~Wright, ``Robust principal component
  analysis?'' \emph{J. ACM}, vol.~58, no.~3, p.~11, May 2011.

\bibitem{Rubinstein:PROC:10}
R.~Rubinstein, A.~M. Bruckstein, and M.~Elad, ``Dictionaries for sparse
  representation modeling,'' \emph{Proc. IEEE}, vol.~98, no.~6, pp. 1045--1057,
  2010.

\bibitem{Lee:NIPS:01}
D.~D. Lee and H.~S. Seung, ``Algorithms for non-negative matrix
  factorization,'' in \emph{Proc. Neural Inform. Process. Syst. Conf.}, 2001,
  pp. 556--562.

\bibitem{Ling:IP:15}
S.~Ling and T.~Strohmer, ``Self-calibration and biconvex compressive sensing,''
  \emph{Inverse Problems}, vol.~31, no.~11, p. 115002, 2015.

\bibitem{Zhu:TSP:11}
H.~Zhu, G.~Leus, and G.~B. Giannakis, ``Sparsity-cognizant total least-squares
  for perturbed compressive sampling,'' \emph{IEEE Trans. Signal Process.},
  vol.~59, no.~5, pp. 2002--2016, May 2011.

\bibitem{Kaleh:TCOM:94}
G.~K. Kaleh and R.~Vallet, ``Joint parameter estimation and symbol detection
  for linear or nonlinear unknown channels,'' \emph{IEEE Trans. Commun.},
  vol.~42, pp. 2406--2413, Jul. 1994.

\bibitem{Candes:TIT:11}
E.~J. Cand{\`e}s and Y.~Plan, ``Tight oracle inequalities for low-rank matrix
  recovery from a minimal number of noisy random measurements,'' \emph{IEEE
  Trans. Inform. Theory}, vol.~57, no.~4, pp. 2342--2359, 2011.

\bibitem{Waters:NIPS:11}
A.~E. Waters, A.~C. Sankaranarayanan, and R.~G. Baraniuk, ``{SpaRCS}:
  {R}ecovering low-rank and sparse matrices from compressive measurements,'' in
  \emph{Proc. Neural Inform. Process. Syst. Conf.}, 2011, pp. 1089--1097.

\bibitem{Lin:10}
Z.~Lin, M.~Chen, L.~Wu, and Y.~Ma, ``The augmented {L}agrange multiplier method
  for exact recovery of corrupted low-rank matrices,'' \emph{arXiv:1009.5055},
  2010.

\bibitem{Wen:MPC:12}
Z.~Wen, W.~Yin, and Y.~Zhang, ``Solving a low-rank factorization model for
  matrix completion by a nonlinear successive over-relaxation algorithm,''
  \emph{Mathematical Programming Computation}, vol.~4, pp. 333--361, 2012.

\bibitem{Balzano:ALL:10}
L.~Balzano, R.~Nowak, and B.~Recht, ``Online identification and tracking of
  subspaces from highly incomplete information,'' in \emph{Proc. Allerton Conf.
  Commun. Control Comput.}, Sep. 2010, pp. 704--711.

\bibitem{Kyrillidis:JMIV:14}
A.~Kyrillidis and V.~Cevher, ``Matrix recipes for hard thresholding methods,''
  \emph{J. Math. Imaging Vis.}, vol.~48, pp. 235--265, 2014.

\bibitem{Babacan:TSP:12}
S.~D. Babacan, M.~Luessi, R.~Molina, and A.~K. Katsaggelos, ``Sparse {B}ayesian
  methods for low-rank matrix estimation,'' \emph{IEEE Trans. Signal Process.},
  vol.~60, no.~8, pp. 3964--3977, Aug. 2012.

\bibitem{He:CVPR:12}
J.~He, L.~Balzano, and A.~Szlam, ``Incremental gradient on the {G}rassmannian
  for online foreground and background separation in subsampled video,'' in
  \emph{Proc. IEEE Conf. Comp. Vision Pattern Recog.}, 2012, pp. 1568--1575.

\bibitem{Aharon:TSP:06}
M.~Aharon, M.~Elad, and A.~Bruckstein, ``{k-SVD: A}n algorithm for designing
  overcomplete dictionaries for sparse representation,'' \emph{IEEE Trans.
  Signal Process.}, vol.~54, no.~11, pp. 4311--4322, 2006.

\bibitem{Mairal:JMLR:10}
J.~Mairal, F.~Bach, J.~Ponce, and G.~Sapiro, ``Online learning for matrix
  factorization and sparse coding,'' \emph{J. Mach. Learn. Res.}, vol.~11, pp.
  19--60, Jan. 2010.

\bibitem{Spielman:COLT:12}
D.~A. Spielman, H.~Wang, and J.~Wright, ``Exact recovery of sparsely-used
  dictionaries,'' in \emph{Proc. Conf. Learning Thy.}, 2012, pp. 37.1--37.18.

\bibitem{Parker:TSP:14a}
J.~T. Parker, P.~Schniter, and V.~Cevher, ``Bilinear generalized approximate
  message passing---{Part I: D}erivation,'' \emph{IEEE Trans. Signal Process.},
  vol.~62, no.~22, pp. 5839--5853, Nov. 2014.

\bibitem{Parker:TSP:14b}
------, ``Bilinear generalized approximate message passing---{Part II:
  A}pplications,'' \emph{IEEE Trans. Signal Process.}, vol.~62, no.~22, pp.
  5854--5867, Nov. 2014.

\bibitem{Kabashima:TIT:16}
Y.~Kabashima, F.~Krzakala, M.~M{\'e}zard, A.~Sakata, and L.~Zdeborov{\'a},
  ``Phase transitions and sample complexity in {B}ayes-optimal matrix
  factorization,'' \emph{IEEE Trans. Inform. Theory}, vol.~62, no.~7, pp.
  4228--4265, 2016.

\bibitem{Donoho:PNAS:09}
D.~L. Donoho, A.~Maleki, and A.~Montanari, ``Message passing algorithms for
  compressed sensing,'' \emph{Proc. Nat. Acad. Sci.}, vol. 106, no.~45, pp.
  18\,914--18\,919, Nov. 2009.

\bibitem{Rangan:ISIT:11}
S.~Rangan, ``Generalized approximate message passing for estimation with random
  linear mixing,'' in \emph{Proc. IEEE Int. Symp. Inform. Thy.}, Aug. 2011, pp.
  2168--2172, (full version at \emph{arXiv:1010.5141}).

\bibitem{Matsushita:NIPS:13}
R.~Matsushita and T.~Tanaka, ``Low-rank matrix reconstruction and clustering
  via approximate message passing,'' in \emph{Proc. Neural Inform. Process.
  Syst. Conf.}, 2013, pp. 917--925.

\bibitem{Lesieur:JSM:17}
T.~Lesieur, F.~Krzakala, and L.~Zdeborová, ``Constrained low-rank matrix
  estimation: {P}hase transitions, approximate message passing and
  applications,'' \emph{J. Stat. Mech.}, vol. 2017, no.~7, p. 073403, 2017.

\bibitem{Miolane:17}
L.~Miolane, ``Fundamental limits of low-rank matrix estimation: {T}he
  non-symmetric case,'' \emph{arXiv:1702.00473}, 2017.

\bibitem{Bilen:TSP:14}
C.~Bilen, G.~Puy, and R.~Gribonval, ``Convex optimization approaches for blind
  sensor calibration using sparsity,'' \emph{IEEE Trans. Signal Process.},
  vol.~62, no.~18, pp. 4847--4856, 2014.

\bibitem{Ahmed:TIT:14}
A.~Ahmed, B.~Recht, and J.~Romberg, ``Blind deconvolution using convex
  programming,'' \emph{IEEE Trans. Inform. Theory}, vol.~60, no.~3, pp.
  1711--1732, 2014.

\bibitem{Hedge:TSP:11}
C.~Hegde and R.~G. Baraniuk, ``Sampling and recovery of pulse streams,''
  \emph{IEEE Trans. Signal Process.}, vol.~59, no.~14, pp. 1505--1517, 2011.

\bibitem{Agarwal:AS:12}
A.~Agarwal, S.~Negahban, and M.~J. Wainwright, ``Matrix decomposition via
  convex relaxation: {O}ptimal rates in high dimensions,'' \emph{Ann.
  Statist.}, vol.~40, no.~2, pp. 1171--1197, 2012.

\bibitem{Wright:II:13}
J.~Wright, A.~Ganesh, K.~Min, and Y.~Ma, ``Compressive principal component
  pursuit,'' \emph{Inform. Inference}, vol.~2, no.~1, pp. 32--68, 2013.

\bibitem{Davenport:JSTSP:16}
M.~A. Davenport and J.~Romberg, ``An overview of low-rank matrix recovery from
  incomplete observations,'' \emph{IEEE J. Sel. Topics Signal Process.},
  vol.~10, no.~4, pp. 608--622, 2016.

\bibitem{Parker:JSTSP:16}
J.~T. Parker and P.~Schniter, ``Parametric bilinear generalized approximate
  message passing,'' \emph{IEEE J. Sel. Topics Signal Process.}, vol.~10,
  no.~4, pp. 795--808, 2016.

\bibitem{Schulke:PRE:16}
C.~Sch{\"u}lke, P.~Schniter, and L.~Zdeborov{\'a}, ``Phase diagram of matrix
  compressed sensing,'' \emph{Physical Rev. E}, vol.~94, no.~6, pp.
  062\,136(1--16), Dec. 2016.

\bibitem{Candes:CPAM:13}
E.~J. Cand{\`e}s, T.~Strohmer, and V.~Voroninski, ``{PhaseLift: Exact} and
  stable signal recovery from magnitude measurements via convex programming,''
  \emph{Commun. Pure \& Appl. Math.}, vol.~66, no.~8, pp. 1241--1274, 2013.

\bibitem{Romanov:PNAS:18}
E.~Romanov and M.~Gavish, ``Near-optimal matrix recovery from random linear
  measurements,'' \emph{Proc. Nat. Acad. Sci.}, 2018.

\bibitem{Metzler:TIT:16}
C.~A. Metzler, A.~Maleki, and R.~G. Baraniuk, ``From denoising to compressed
  sensing,'' \emph{IEEE Trans. Inform. Theory}, vol.~62, no.~9, pp. 5117--5144,
  2016.

\bibitem{Berthier:17}
R.~Berthier, A.~Montanari, and P.-M. Nguyen, ``State evolution for approximate
  message passing with non-separable functions,'' \emph{arXiv:1708.03950},
  2017.

\bibitem{Rush:ISIT:16}
C.~Rush and R.~Venkataramanan, ``Finite-sample analysis of approximate message
  passing,'' in \emph{Proc. IEEE Int. Symp. Inform. Thy.}, 2016, pp. 755--759.

\bibitem{Bayati:TIT:11}
M.~Bayati and A.~Montanari, ``The dynamics of message passing on dense graphs,
  with applications to compressed sensing,'' \emph{IEEE Trans. Inform. Theory},
  vol.~57, no.~2, pp. 764--785, Feb. 2011.

\bibitem{Bayati:AAP:15}
M.~Bayati, M.~Lelarge, and A.~Montanari, ``Universality in polytope phase
  transitions and message passing algorithms,'' \emph{Ann. App. Prob.},
  vol.~25, no.~2, pp. 753--822, 2015.

\bibitem{Rangan:TIT:16a}
S.~Rangan, P.~Schniter, E.~Riegler, A.~Fletcher, and V.~Cevher, ``Fixed points
  of generalized approximate message passing with arbitrary matrices,''
  \emph{IEEE Trans. Inform. Theory}, vol.~62, no.~12, pp. 7464--7474, Dec.
  2016.

\bibitem{Caltagirone:ISIT:14}
F.~Caltagirone, F.~Krzakala, and L.~Zdeborov\'a, ``On convergence of
  approximate message passing,'' in \emph{Proc. IEEE Int. Symp. Inform. Thy.},
  Jul. 2014, pp. 1812--1816.

\bibitem{Vila:ICASSP:15}
J.~Vila, P.~Schniter, S.~Rangan, F.~Krzakala, and L.~Zdeborov{\'a}, ``Adaptive
  damping and mean removal for the generalized approximate message passing
  algorithm,'' in \emph{Proc. IEEE Int. Conf. Acoust. Speech \& Signal
  Process.}, 2015, pp. 2021--2025.

\bibitem{Rangan:VAMP}
S.~Rangan, P.~Schniter, and A.~K. Fletcher, ``Vector approximate message
  passing,'' \emph{arXiv:1610.03082}, 2016.

\bibitem{Opper:NIPS:05}
M.~Opper and O.~Winther, ``Expectation consistent free energies for approximate
  inference,'' in \emph{Proc. Neural Inform. Process. Syst. Conf.}, 2005, pp.
  1001--1008.

\bibitem{Fletcher:ISIT:16}
A.~K. Fletcher, M.~Sahraee-Ardakan, S.~Rangan, and P.~Schniter, ``Expectation
  consistent approximate inference: {G}eneralizations and convergence,'' in
  \emph{Proc. IEEE Int. Symp. Inform. Thy.}, 2016, pp. 190--194.

\bibitem{Ma:IA:17}
J.~Ma and L.~Ping, ``Orthogonal {AMP},'' \emph{IEEE Access}, vol.~5, pp.
  2020--2033, 2017.

\bibitem{Takeuchi:ISIT:17}
K.~Takeuchi, ``Rigorous dynamics of expectation-propagation-based signal
  recovery from unitarily invariant measurements,'' in \emph{Proc. IEEE Int.
  Symp. Inform. Thy.}, 2017, pp. 501--505.

\bibitem{Fletcher:NIPS:18}
A.~K. Fletcher, S.~Rangan, S.~Sarkar, and P.~Schniter, ``Plug-in estimation in
  high-dimensional linear inverse problems: {A} rigorous analysis,'' in
  \emph{Proc. Neural Inform. Process. Syst. Conf.}, 2018, (see also
  \emph{arXiv:1806.10466}).

\bibitem{Fletcher:NIPS:17}
A.~K. Fletcher, M.~Sahraee-Ardakan, S.~Rangan, and P.~Schniter, ``Rigorous
  dynamics and consistent estimation in arbitrarily conditioned linear
  systems,'' in \emph{Proc. Neural Inform. Process. Syst. Conf.}, 2017, pp.
  2542--2551.

\bibitem{Wainwright:FTML:08}
M.~J. Wainwright and M.~I. Jordan, ``Graphical models, exponential families,
  and variational inference,'' \emph{Found. Trends Mach. Learn.}, vol.~1, May
  2008.

\bibitem{Minka:Diss:01}
T.~Minka, ``A family of approximate algorithms for {B}ayesian inference,''
  Ph.D. dissertation, Dept. Comp. Sci. Eng., MIT, Cambridge, MA, Jan. 2001.

\bibitem{Seeger:Tech:05}
M.~Seeger, ``Expectation propagation for exponential families,'' EPFL, Tech.
  Rep. 161464, 2005.

\bibitem{Heskes:UAI:02}
T.~Heskes and O.~Zoeter, ``Expectation propagation for approximate inference in
  dynamic {B}ayesian networks,'' in \emph{Proc. Uncertainty Artif. Intell.},
  2002, pp. 313--320.

\bibitem{Tulino:TIT:13}
A.~M. Tulino, G.~Caire, S.~Verd{\'u}, and S.~{Shamai (Shitz)}, ``Support
  recovery with sparsely sampled free random matrices,'' \emph{IEEE Trans.
  Inform. Theory}, vol.~59, no.~7, pp. 4243--4271, Jul. 2013.

\bibitem{Reeves:ALL:17}
G.~Reeves, ``Additivity of information in multilayer networks via additive
  {G}aussian noise transforms,'' in \emph{Proc. Allerton Conf. Commun. Control
  Comput.}, 2017, pp. 1064--1070.

\bibitem{Reeves:ISIT:16}
G.~Reeves and H.~D. Pfister, ``The replica-symmetric prediction for compressed
  sensing with {G}aussian matrices is exact,'' in \emph{Proc. IEEE Int. Symp.
  Inform. Thy.}, 2016.

\bibitem{Barbier:ALL:16}
J.~Barbier, M.~Dia, N.~Macris, and F.~Krzakala, ``The mutual information in
  random linear estimation,'' in \emph{Proc. Allerton Conf. Commun. Control
  Comput.}, 2016, pp. 625--632.

\bibitem{Dempster:JRSS:77}
A.~Dempster, N.~M. Laird, and D.~B. Rubin, ``Maximum-likelihood from incomplete
  data via the {E}{M} algorithm,'' \emph{J. Roy. Statist. Soc.}, vol.~39, pp.
  1--17, 1977.

\bibitem{Neal:Jordan:98}
R.~Neal and G.~Hinton, ``A view of the {EM} algorithm that justifies
  incremental, sparse, and other variants,'' in \emph{Learning in Graphical
  Models}, M.~I. Jordan, Ed.\hskip 1em plus 0.5em minus 0.4em\relax MIT Press,
  1998, pp. 355--368.

\bibitem{Kamilov:TIT:14}
U.~S. Kamilov, S.~Rangan, A.~K. Fletcher, and M.~Unser, ``Approximate message
  passing with consistent parameter estimation and applications to sparse
  learning,'' \emph{IEEE Trans. Inform. Theory}, vol.~60, no.~5, pp.
  2969--2985, May 2014.

\bibitem{Vila:TSP:13}
J.~P. Vila and P.~Schniter, ``Expectation-maximization {G}aussian-mixture
  approximate message passing,'' \emph{IEEE Trans. Signal Process.}, vol.~61,
  no.~19, pp. 4658--4672, Oct. 2013.

\bibitem{Dabov:TIP:07}
K.~Dabov, A.~Foi, V.~Katkovnik, and K.~Egiazarian, ``Image denoising by sparse
  {3-D} transform-domain collaborative filtering,'' \emph{IEEE Trans. Image
  Process.}, vol.~16, no.~8, pp. 2080--2095, 2007.

\bibitem{Rudin:PhyD:92}
L.~I. Rudin, S.~Osher, and E.~Fatemi, ``Nonlinear total variation based noise
  removal algorithms,'' \emph{Physica D}, vol.~60, pp. 259--268, 1992.

\bibitem{Beck:TIP:09}
A.~Beck and M.~Teboulle, ``Fast gradient-based algorithms for constrained total
  variation image denoising and deblurring problem,'' \emph{IEEE Trans. Image
  Process.}, vol.~18, no.~11, pp. 2419--2434, 2009.

\bibitem{Romano:JIS:17}
Y.~Romano, M.~Elad, and P.~Milanfar, ``The little engine that could:
  {R}egularization by denoising {(RED)},'' \emph{SIAM J. Imag. Sci.}, vol.~10,
  no.~4, pp. 1804--1844, 2017.

\bibitem{Reehorst:TCI:19}
E.~T. Reehorst and P.~Schniter, ``Regularization by denoising: {C}larifications
  and new interpretations,'' \emph{IEEE Trans. Comp. Imag.}, vol.~5, no.~1, pp.
  52--67, Mar. 2019.

\bibitem{Goldstein:14}
T.~Goldstein, C.~Studer, and R.~Baraniuk, ``Forward-backward splitting with a
  {FASTA} implementation,'' \emph{arXiv:1411.3406}, 2014.

\end{thebibliography}

\end{document}